\documentclass[12pt]{article} 

\usepackage[utf8]{inputenc}
\usepackage[english]{babel}
\usepackage{graphicx}
\graphicspath{ {images/} }
\usepackage[a4paper,right=1in,left=1in,top=1in,bottom=1in]{geometry}
\usepackage{fancyhdr}
\usepackage{rotating}
\usepackage{placeins}

\usepackage[round]{natbib}

\usepackage[dvipsnames]{xcolor}
\definecolor{light-gray}{gray}{0.95}
\colorlet{laurencol}{cyan!10!light-gray}
\colorlet{laurencolvert}{cyan!25!light-gray}
\colorlet{laurencol2}{red!10!light-gray}
\usepackage{colortbl}
\usepackage{arydshln}
\usepackage{enumitem}
\usepackage{setspace}

\makeatletter
\newcommand\notsotiny{\@setfontsize\notsotiny\@vipt\@viipt}
\makeatother

\usepackage{tikz}

\usepackage{tikz}
\usetikzlibrary{shapes,arrows,decorations.markings}
\usetikzlibrary{positioning}

\usepackage{amsmath, amsfonts, amssymb} 
\usepackage[labelfont=bf]{caption}
\usepackage{setspace}
\usepackage{accents} 
\usepackage{color, soul}
\usepackage[title,titletoc,toc]{appendix}
\usepackage{rotating} 
\usepackage{float} 
\usepackage{multirow} 
\usepackage[position=top]{subfig}
\usepackage[subfigure]{tocloft}
\usepackage{tikz}
\usetikzlibrary{decorations.pathreplacing,calc}
\usepackage[many]{tcolorbox}
\newtcolorbox{rightbrace}{%
    enhanced jigsaw, 
    breakable, 
    frame hidden, 
    parbox=false,
}
\usepackage[explicit]{titlesec}
\graphicspath{ {images/} }

\allowdisplaybreaks
\usepackage{color}   
\usepackage{hyperref}
\hypersetup{
    colorlinks=true, 
    linktoc=all,     
    linkcolor=black,  
    filecolor=black,
    citecolor = black,      
    urlcolor=black,
}

\newcommand{\appropto}{\mathrel{\vcenter{
  \offinterlineskip\halign{\hfil$##$\cr
    \propto\cr\noalign{\kern2pt}\sim\cr\noalign{\kern-2pt}}}}}

\title{Statistical methods for partitioning ribbon and globally-distributed flux using data from the Interstellar Boundary Explorer}
\author{Lauren J. Beesley$^{*1,2}$, Dave Osthus$^2$, Kelly R. Moran$^2$, Madeline A. Stricklin$^2$, \\
Grant David Meadors$^3$, Thomas K. Kim$^4$, Sung Jun Noh$^4$, Nehpreet K. Walia$^4$, \\
Paul H. Janzen$^4$, Eric J. Zirnstein$^6$, Brian P. Weaver$^2$, Daniel B. Reisenfeld$^4$\\
\\
$^1$Information Systems \& Modeling, Los Alamos National Laboratory\\
$^2$Statistical Sciences, Los Alamos National Laboratory\\
$^3$Space Remote Sensing and Data Science, Los Alamos National Laboratory\\
$^4$Space Science and Applications, Los Alamos National Laboratory\\
$^5$Physics and Astronomy, University of Montana\\
$^6$Astrophysical Sciences, Princeton University\\
\\
*Corresponding Author: lvandervort@lanl.gov
}
\date{}
\begin{document}

\maketitle

 \doublespacing
 
\begin{abstract}
NASA’s Interstellar Boundary Explorer (IBEX) satellite collects data on energetic neutral atoms (ENAs) that can provide insight into the heliosphere boundary between our solar system and interstellar space. Using these data, scientists can construct \textit{maps} of the ENA intensities (often, expressed in terms of flux) observed in all directions. The ENA flux observed in these maps is believed to come from at least two distinct sources: one source which manifests as a \textit{ribbon} of concentrated ENA flux and one source (or possibly several) that results in a smoothly-varying globally-distributed flux. Each ENA source type and its corresponding ENA intensity map is of separate scientific interest. \\
\indent In this paper, we develop statistical algorithms for separating the total ENA intensity maps into two source-specific maps (ribbon and globally-distributed flux) and estimating corresponding uncertainty. Key advantages of the proposed method include enhanced model flexibility and improved propagation of estimation uncertainty. We evaluate the proposed methods on simulated data designed to mimic realistic data settings. We also propose new methods for estimating the center of the near-elliptical ribbon in the sky, which can be used in the future to study the location and variation of the local interstellar magnetic field. 
\end{abstract}

\section{Introduction}
\begin{figure}[h!]
  \centering
\includegraphics[trim={0cm 3.5cm 0cm 3cm}, clip, width=6in]{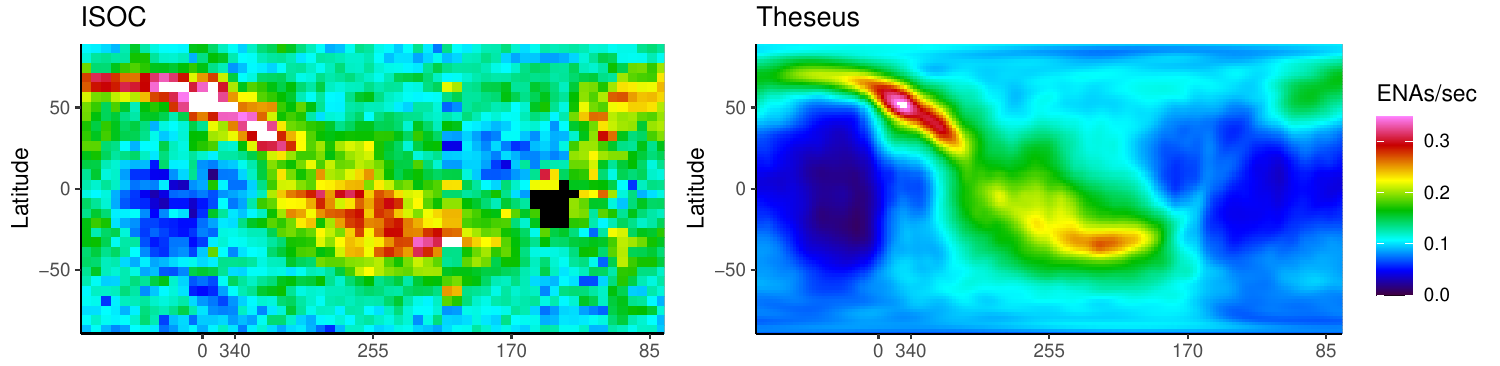}
\caption{ENA rate maps generated by the IBEX Science Operations Center (left) and by \citet{Osthus2022} (right) using IBEX data collected in 2010. \\
\footnotesize The ISOC map (left) corresponds to the entire year of 2010 using data observed in the ram direction with additional data corrections applied. The Theseus map by \citet{Osthus2022} (right) corresponds to the first 6 months of 2010 and incorporates data collected in the ram and antiram directions. Data are plotted in terms of ENAs/sec.  }
\label{2010_data}
\vspace{-0.5cm}
\end{figure}

The Interstellar Boundary Explorer (IBEX) satellite is part of an Earth-orbiting U.S. National Aeronautics and Space Administration (NASA) mission to study the boundaries of the bubble-like region of space called the heliosphere surrounding our solar system \citep{McComas2009}. IBEX accomplishes this with two instruments, IBEX-Hi and IBEX-Lo, which observe energetic neutral atoms (ENAs) that emanate from the heliosheath, the outer layer of the heliosphere. Here, we focus on higher-energy ENAs detected by the IBEX-Hi imager \citep{Funsten2009}. These ENAs are primarily fast-moving ($\sim$100 - 1000 km/s) neutral hydrogen atoms that originate from solar wind protons that charge exchange with incoming interstellar neutral hydrogen and are then redirected back sunward. Once a charged particle becomes an ENA, it continues moving on a ballistic trajectory, undeflected by the magnetic field that permeates the heliosphere. Some of these particles travel toward Earth and are measured by the IBEX-Hi imager. \\
\indent The IBEX spacecraft spins about a central axis that is redirected toward the Sun every few days. As the spacecraft spins, the IBEX-Hi instrument views a circular slice of sky perpendicular to its spin axis. As the Earth orbits the Sun carrying IBEX with it, IBEX-Hi is able to image the entire sky over the course of six months. When pointing in a particular direction, IBEX-Hi observes ENAs coming from a range of directions about the central line-of-sight of the instrument, with the majority of the distribution within roughly 6.5 degrees of the center of the field of view \citep{Schwadron2009}. This wide field of view induces spatial blurring of the ENA arrival directions. The rates at which ENAs are observed by IBEX-Hi as a function of calendar time, latitude/longitude, and particle energy can inform our understanding of particle dynamics within the heliosphere \citep{Zirnstein2019}, can be used to infer the direction of the interstellar magnetic field near the heliosphere \citep{Zirnstein2016b}, and can even be used to construct 3-dimensional estimates of heliosphere structure \citep{Reisenfeld2021}.\\
\indent In service of these objectives, scientists use IBEX ENA data to construct pixelated \textit{maps} of ENA rates/intensities (i.e., ENAs per second) and uncertainties as a function of latitude and longitude. Sometimes, these maps are expressed in terms of differential flux, i.e., the ENAs per squared centimeter, per second, per steradian, per keV (kilo-electron volt); we will treat ENA rates and fluxes interchangeably in this work. These ENA rate or flux maps are generated separately for ENAs falling in different overlapping energy steps or bins, where energy of incoming ENAs is approximated by an onboard electrostatic analyzer (ESA) \citep{Funsten2009}. Throughout, we will refer to these energy steps as ``ESAs" after the equipment used to bin ENA energies. The lowest trusted energy step measured by IBEX, ESA 2, corresponds to $\sim$0.7 keV energy particles, and the highest energy step, ESA 6, corresponds to $\sim$4.3 keV particles. New maps are generated yearly or biyearly throughout the mission. Maps are also generated separately for ENAs collected looking in the ram and antiram directions, where the ``ram" direction refers to the direction the Earth is moving around the Sun and where ``antiram" refers to ENAs coming from the opposite direction. Historically, these maps were all generated and publicly released by the IBEX Science Operations Center (ISOC) as ecliptic coordinate frame maps (a celestial coordinate system commonly used to represent the orientation of objects in our solar system) composed of 6 degree longitude by 6 degree latitude pixels, and recently-published and statistically rigorous ``Theseus" maps provide finer pixel resolution maps composed of 2 x 2 degree pixels \citep{Schwadron2009, Osthus2022}. In addition to higher resolution and more rigorous methodology, the ``Theseus" map-making procedure also addresses the spatial blurring of the IBEX-Hi instrument viewing aperture through deconvolution, together providing an opportunity for new, exciting advances in heliospheric science.  \\
\indent \textbf{Figure \ref{2010_data}} shows example ISOC and Theseus ENA maps generated using data observed in 2010. In both maps, we observe regions of ENA enhancements in the center and at the left and right sides. These enhancements are a consequence of the direction of motion of the Sun and heliosphere through interstellar space. Imagining the heliosphere as a boat moving through water, the enhancement at the center of the map represents the nose of the boat and the enhancements on the left and right sides represent the back or tail of the boat. Also evident in these maps is a band of higher ENA intensity cutting across the map. Called the IBEX ``ribbon", this band of ENA enhancement was completely unanticipated by heliospheric scientists prior to the IBEX mission. \\
\indent The discovery of the IBEX ribbon was lauded on the cover of \textit{Science} and ignited a decade of research into the physical underpinnings of ribbon formation that continues today \citep{McComas2009}. Scientists now believe that heliospheric ENAs come from at least two distinct sources within the heliosheath, which is divided into inner and outer components delineated by a boundary called the heliopause. These two sources are called (1) the globally-distributed flux (GDF), believed to be primarily generated in the inner heliosheath, and (2) the ENA ribbon, attributed to solar wind protons that are neutralized on their outward trajectory, travel beyond the heliopause, become ionized and trapped in the nearby interstellar magnetic field, and then again are neutralized and return inside the heliopause \citep{McComas2009, Schwadron2014, Zirnstein2015}. The physics behind the ENA ribbon are not fully understood, and there are numerous competing hypotheses \citep{McComas2014}. Both the ribbon and the GDF are independently interesting and are associated with distinct lines of research; maps containing GDF and ribbon together have limited utility for heliospheric science. An important data objective, therefore, is to accurately partition the ENA intensity maps into GDF- and ribbon-specific maps, which can then each be used for subsequent study. \\
\indent Map partitioning is extremely challenging, because the partitioning is not identified by the data alone and must be strongly informed by assumptions. To make matters worse, the heliospheric nose and tail regions of interest for GDF science are both in part within the ribbon's path, so modeling choices used to partition the maps can have a substantial impact on subsequent physics insights in those regions. Additionally, different hypotheses about ribbon formation generally result in subtle differences in the ribbon shape, so assumptions made about the shape will be consequential for discriminating between competing theories. To summarize, the assumptions made when partitioning the ENA ribbon will be consequential and untestable, and they should be general enough and implemented with finesse to avoid obscuring real physical phenomena and introducing bias in subsequent analyses. Despite these very real challenges, the problem is not hopeless. Prior work indicates that the ribbon follows a near-circular (slightly elliptical) path in cartesian space \citep{Funsten2013}. Additionally, the ENA ribbon is believed to change more quickly in space than the underlying GDF. Both of these attributes, combined with physics insights accumulated over the last 14 years of IBEX research, provide a path toward accurate partitioning/separation of the ENA ribbon from the GDF. \\
\indent Unsurprisingly, several authors have proposed strategies for partitioning ENA maps, leveraging these and other assumptions about the ribbon. Due to its near-circularity, the ribbon manifests as a near-horizontal stripe of concentrated ENA emissions when viewed in a particular rotational frame and plotted as a latitude/longitude rectangle in a standard cylindrical projection map (see \textbf{Figure \ref{ribbonsepdiagram}}). \citet{Schwadron2011}, \citet{Dayeh2019}, and \citet{Reisenfeld2021} use this rotational frame to facilitate ribbon separation, where parametric modeling of the height and shape of ribbon cross-sections (i.e., vertical slices in the rotated frame) is used to obtain the partitioning. These existing approaches have many limitations. Firstly, a common assumption is that ribbon cross-sections mirror Gaussian densities. This assumption is restrictive. For example, one of the popular theories of ribbon generation is associated with a distinctive skewed ribbon \citep{Zirnstein2019}, and separations generated assuming the Gaussian-type ribbon cross-sections may irreparably bias subsequent hypothesis evaluations. The ideal map partitioning algorithm should be flexible enough to capture these scientifically-relevant features, facilitating discrimination between competing scientific theories. Secondly, uncertainty in the separated ENA rate maps should be a function of the uncertainty of the total (ribbon + GDF) ENA map and uncertainty due to the separation itself. None of these existing methods account for this latter and substantial source of uncertainty. Recently, \citet{Swaczyna2022} proposed a separation method based on spherical harmonic decompositions that appears to address these two limitations. However, this approach tends to introduce rippling artifacts into the separated maps, it tends to place all lack of fit and model noise into the ribbon map estimate, and it does not leverage any spatial structure present in the ribbon region for GDF map estimation. Thirdly, all of the existing ribbon separation methods were developed and tested using 6-degree observed maps alone; \textit{none} of the existing ribbon separation methods have been validated using simulated maps, directly evaluated against each other, or tested on real data maps with higher than 6-degree resolution. Given the consequential effects of ribbon separation on heliospheric science, additional work must be done before GDF-only and ribbon-only maps meet a scientific standard high enough to be trusted in subsequent analyses of fine-scale (in some cases, even large-scale) GDF and ribbon features.  \\
\indent In this paper, we propose a flexible algorithm for separating the total ENA rate maps into source-specific maps and for estimating corresponding uncertainty. This approach addresses many of the limitations of the existing methods, leverages physics knowledge about the ribbon gained over the last 14 years of IBEX research, and adds a variety of guardrails to improve the robustness and trustworthiness of the ribbon/GDF partitioning. Through a rigorous validation of these methods in simulated maps and against a competing approach in the literature, we demonstrate that the proposed methods perform well in settings likely encountered in real IBEX data. We also highlight the performance of the proposed methods in real Theseus and ISOC IBEX data maps; a detailed presentation of these real data results will be provided in a series of follow-up manuscripts. Auxiliary to the goal of ribbon separation, we also propose new methods for estimating the center of the near-circular ribbon in the sky, which can provide insight into the location and variation of the local interstellar magnetic field lines \citep{Zirnstein2016b}. Together, this work represents a substantial advancement in the processing of IBEX data and illustrates the value of close collaboration between statisticians and scientific domain experts.\\
\indent In \textbf{Section \ref{methods}}, we propose novel methods for ribbon separation and estimation of the ribbon center. We demonstrate the performance of these methods in simulations in \textbf{Section \ref{simresults}} in real IBEX data in \textbf{Section \ref{mainrealdata}}. In \textbf{Section \ref{discuss}}, we provide some discussion.

\section{Methods} \label{methods}
\begin{figure}[h!]
  \centering
\includegraphics[trim={0cm 0cm 0cm 0cm}, clip, width=6in]{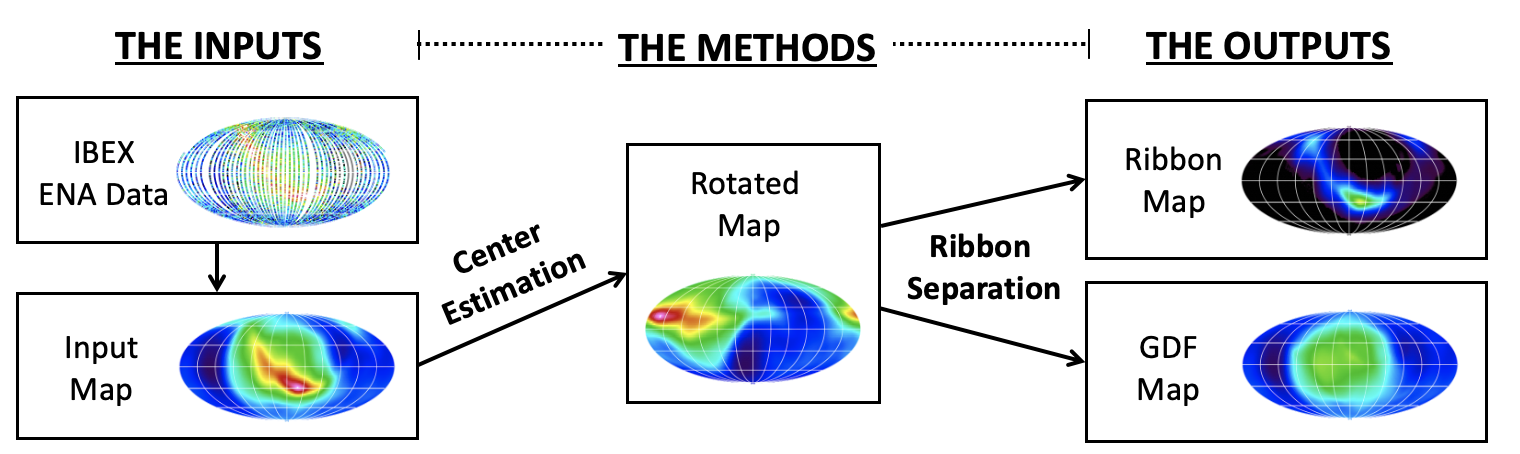}
\caption{Visualization of IBEX data analysis pipeline and role of proposed methods}
\label{pipeline}
\end{figure}

IBEX ``maps" are pixelated representations of the intensity of ENAs observed from the IBEX satellite in all directions (e.g., \citet{Funsten2009b}). Estimated maps are generated through a map-making pipeline that converts data on encounters with ENAs into an estimated ENA rate or differential flux and corresponding uncertainty for each of many spatial latitude/longitude bins of the sky. Traditionally, the IBEX Science Operations Center (ISOC) generates 6 degree maps using simple method-of-moments estimators and ad hoc nearest neighbor spatial smoothing (\url{https://ibex.princeton.edu/DataRelease}). Recently, \citet{Osthus2022} developed a higher-resolution and statistically rigorous map-making method involving spatial modeling of the ENA rate data and deconvolution to address instrument-related spatial blurring. The methods proposed in our current paper can be applied to either type of input map, and we will demonstrate our methods both using 2-degree ``Theseus" maps generated by \citet{Osthus2022} and using 6-degree maps generated by the IBEX Science Operations Center (ISOC). \\
\indent We assume that this map-making process has already been implemented, outputting a vector $\hat D$ of length $J \times K$ representing the estimated ENA rate for each of $J$ longitude bins and $K$ latitude bins of the sky along with an estimated $J \times K$ vector of corresponding uncertainty estimates, $\hat \sigma_D^2$. The map-making process may also produce a $J \times K$ by $J \times K$ covariance matrix $\hat \Sigma_D$ associated with estimated $\hat D$. These pixelated maps and their uncertainties are generated in the ecliptic coordinate system, a celestial coordinate system commonly used to represent the orientation of objects in our solar system.\\
\indent In this section, we describe the proposed methods for addressing the following challenges: (1) estimating the spatial ``center" of the ENA ribbon and (2) separating the ENA map into ribbon-only and GDF-only maps and estimating corresponding uncertainties. \textbf{Supp. Section A} details our strategy for rotating pixelated maps between different spherical reference frames. The IBEX-Hi data pipeline is summarized in \textbf{Figure \ref{pipeline}}. 

\subsection{Estimating the ribbon center} \label{center}
The ribbon center roughly refers to an axis of symmetry passing through the ribbon in three-dimensional space. In this section, we propose a strategy for estimating the ribbon center and corresponding uncertainty using input map $\hat D$ and either covariance $\hat \Sigma_D$ or its diagonal, $\hat \sigma_D^2$. First, we clarify what we mean by the ribbon ``center." Suppose we view the ENA data collected in latitude/longitude space in terms of a unit sphere. We can visualize the ENA ribbon as being a closed shape on the surface of the sphere organized nearly-symmetrically around a normal axis/line intersecting the sphere. We will formalize the ribbon ``center" as the latitude and longitude in which this normal line intersects the sphere. Strictly speaking, this line will intersect the unit sphere in two places, and we will define the ribbon ``center" as the point that lies closer in latitude and longitude to the ribbon. \\
\indent Two complications to estimating this center are that (1) the ``location" of the ribbon is not well-defined and (2) the ENA map $\hat D$ used for estimation is pixelated and estimated with error. To tackle the former problem, we build on existing work that defines the ``location" of the ribbon on the sphere in terms of estimated ribbon peaks. In \textbf{Supp. Section D}, we also provide an alternative summary of the ribbon location. We address the latter problem of map pixelation by repeating our center estimation procedure multiple times on ENA maps drawn from a multivariate normal distribution as detailed in \textbf{Supp. Section G.1}.  

  \begin{figure}[h!]
  \centering
\includegraphics[trim={0cm 0cm 0cm 0cm}, clip, width=6in]{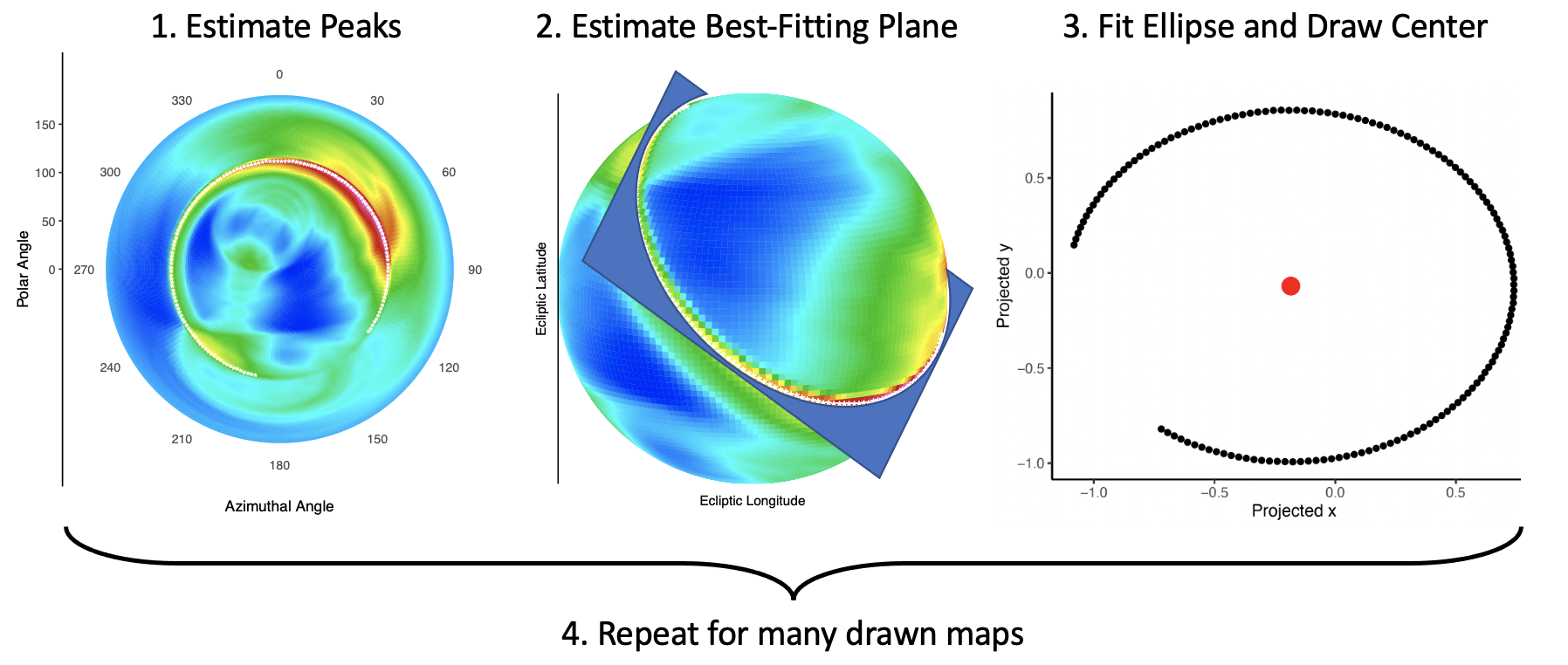}
\caption{Visualization of proposed ribbon center estimation method, illustrated for a simulated GDF + ribbon weak scattering map. These methods can also be applied to ribbon-only map estimates.\\
 \footnotesize The ribbon center estimation method is illustrated for a simulated ENA rate map as described in \textbf{Section \ref{simresults}} under assumptions of weak scattering. The first panel shows the map plotted in polar coordinates after rotation, where the ribbon center is located at polar angle 180. The ribbon is visible as a near-circular band of enhanced ENA intensity, and the estimated ribbon peaks are shown as small white dots along the ribbon. The second panel illustrates the best fitting plane passing through these peak points in three dimensions. The third panel illustrates these peak points projected onto the best fitting plane and the corresponding estimated ellipse center (red dot). This ellipse center is converted to a latitude/longitude on the surface of the unit sphere by calculating the intersection between unit sphere and a line passing through the estimated ellipse center (in the plane) that is perpendicular to the best fitting plane. In the third panel, this perpendicular line would point out of the page.  }
 \label{centerdiagram}
\end{figure}
\indent \textbf{Figure \ref{centerdiagram}} illustrates the proposed ribbon center estimation procedure. After drawing a realization of the input map from a multivariate normal distribution accounting for its uncertainty, we rotate the drawn map into a rotational frame where an assumed working value for the ribbon center sits at the North pole (polar angle 180). We then:
\begin{enumerate}[noitemsep, label={\arabic*}.]
    \item Estimate the ribbon peak location for each azimuthal (i.e., longitude) sector in the rotated frame using cubic spline interpolation. Update peak locations using local scatterplot smoothing and outlier pruning across azimuthal sectors. 
     \item Estimate the best fitting plane through these peak points using singular value decomposition as described in \citet{Blum2015}. Project the peak points onto the best fitting plane using Rodrigues vector rotation \citep{Dai2015}.
    \item Fit an ellipse using the \textit{fitConic} package in R \citep{Fitzgibbon1999}. Then, conservatively estimate the uncertainty in the ribbon center for the given map draw as the variance of the ellipse fit residuals.
    \item Finally, draw Cartesian coordinates for the ellipse center in the plane from a bivariate normal distribution and calculate the corresponding latitude and longitude on the unit sphere (i.e., the ribbon center) in which the unit sphere intersects a normal line passing through the ellipse center in the best fitting plane.  
\end{enumerate}
This procedure is repeated to obtain many samples (e.g., 500) for the ribbon center latitude and longitude in ecliptic coordinates. The means of these samples provide our overall ribbon center location estimate. We recommend iterating this process several times, each time using the estimated ribbon center to define the new working rotational frame. Additional details are provided in \textbf{Supp. Section G.1}. \\
\indent Later on, we compare the proposed ribbon center estimation algorithm with the method in \citet{Funsten2013} as detailed in \textbf{Supp. Section G.1}. Key differences from the proposed approach are that \citet{Funsten2013} (1) assumes a Gaussian-type shape for the ribbon cross-sections, (2) ignores uncertainty in the input map, and (3) performs ellipse fitting without iteration all in the working rotational frame.

\subsection{Ribbon separation} \label{separation}
\begin{figure}[h!]
  \centering
\includegraphics[trim={0cm 0cm 0cm 0cm}, clip, width=6.2in]{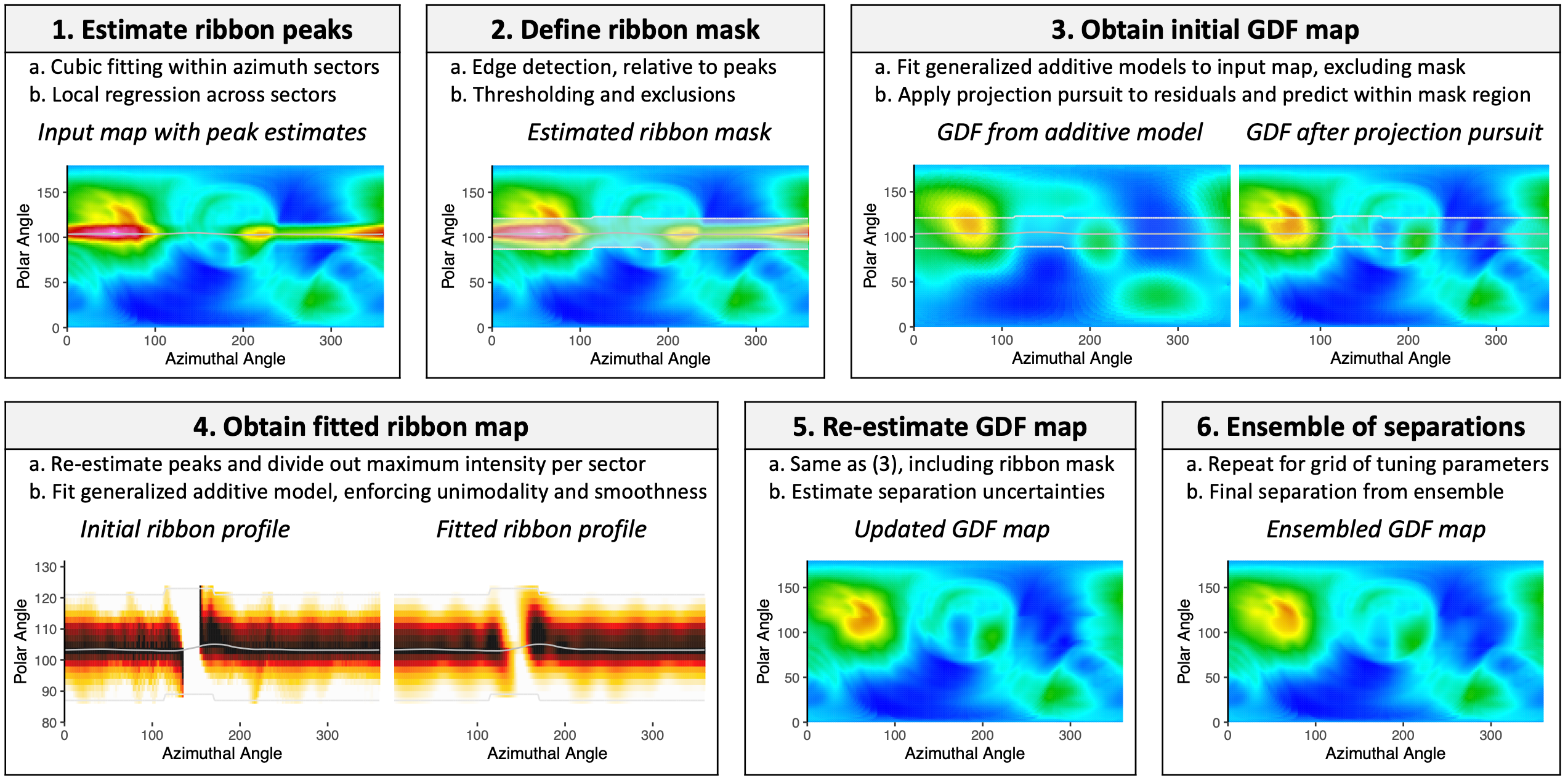}\\
\caption{llustration of the ribbon separation methods for a simulated input map (weak scattering model).\\
 \footnotesize The ribbon separation method is applied to a simulated ENA rate map as described in \textbf{Section \ref{simresults}} under assumptions of weak scattering. Images for steps 1-3 and 5 correspond to ENA rates between 0 (black) and 0.3 (red/magenta) ENAs per second. Images for step 4 correspond to scaled ENA profile intensities between 0 (white) and 1 (black). The ribbon mask is illustrated using a shaded horizontal band in images for steps 2-4. Estimated ribbon peaks are also shown as a near-horizontal gray line in images for steps 1-4.}
 \label{ribbonsepdiagram}
\end{figure}

\indent In this section, we describe our ribbon separation algorithm. Let $\hat M$ denote the vectorized input map estimates and $\sigma^2_M$ denote the uncertainties, \textit{after rotation} into the rotational frame defined by our estimated ribbon center. We assume input map $\hat M$ is the sum of a ribbon-only map, $\hat R$, and a GDF-only map, $\hat G$. The problem of ribbon separation can be broken down into two questions: (1) where do the ribbon and GDF coexist (i.e., where is  $\hat R$ nonzero) and (2) what is the GDF-related ENA rate when both are present (i.e., what is $\hat G$ ``under" the ribbon)? To answer the first question, we apply image processing techniques to identify a block of pixels possibly containing both ribbon and GDF ENAs, collectively called the ribbon \textit{mask}. To answer the second question, we combine various spatial regression techniques to estimate the GDF shape within the ribbon mask region, leveraging physical assumptions. Standard errors, which propagate uncertainty in the input map along with additional uncertainty due to the separation itself, are calculated by repeating the proposed method across bootstrap samples of drawn realizations of the input map. The ``final" separation defined for a given input map is obtained from an ensemble of many candidate separations across different sets of tuning parameter values, where the weight given to each candidate separation is based on a proposed ``goodness of separation" heuristic. \textbf{Figure \ref{ribbonsepdiagram}} illustrates the proposed ribbon separation method for an example simulated input map and a fixed set of tuning parameters. Below, we summarize the assumptions and strategies used for each step of the estimation procedure. Details, including pseudocode, are provided in \textbf{Supp. Section G.2}. 

\subsubsection{Estimating ribbon peaks and defining the ribbon mask} \label{peaksmask}
The goal of ribbon mask estimation is to identify a region in which ribbon-related ENA rates may be non-negligible. A very conservative mask will contain the entire ribbon but will leave little data outside the masked region for GDF estimation, leading to sub-optimal partitioning. Conversely, a less conservative mask may preserve more data for GDF map estimation while failing to capture the entire ribbon. Therefore, care must be taken in defining a ribbon mask region to balance these two opposing objectives. \\
\indent Suppose we visualize the data in the ribbon center rotational frame as a heatmap with latitude/polar angle on the y-axis, longitude/azimuthal angle on the x-axis, and the ENA rate $\hat M$ captured by the color intensity as in \textbf{Figure \ref{ribbonsepdiagram}}. The ENA ribbon manifests as a near-horizontal band of concentrated ENA emissions. To capture deviations from a perfectly horizontal ribbon band, we estimated ribbon peak locations using a minor modification of the method in \textbf{Section \ref{center}} to stabilize estimation for noisy real data. \\
\indent Fixing these estimated ribbon peak locations, we then used a data-driven strategy to define a ribbon mask region, by estimating windows of polar angles above and below the ribbon peaks likely to contain ribbon ENAs. One strategy for identifying the location of the ENA ribbon is to estimate how quickly the ENA emissions are changing as a function of polar angle, because the ENA ribbon is expected to vary more quickly (i.e., have higher spatial gradients) than the GDF \citep{Funsten2013}. Regions associated with changing behavior of spatial gradients are sometimes called edges, and many methods have been developed for identifying these edges on fixed spatial grids. We apply this same logic to identify map regions associated with the ``foothills" of the ribbon ``mountain range." \\
\indent Re-organizing the ENA rate map with longitude on the x-axis and distance from peak (rounded into pixels) on the y-axis, we applied a popular edge detection method called the Sobel operator \citep{kanopoulos1988design}, where higher Sobel values are associated with greater ``edginess." The ribbon mask was defined in terms of the narrowest window that captures all of the top $u^{th}$ percentile of gradient estimates within $v$ degrees of the ribbon peak, with an additional one or two pixels added on either side to ensure the entire ribbon was captured. This process and all subsequent ribbon separation steps were repeated multiple times across a grid of reasonable values for $u$ and $v$.

\subsubsection{Obtaining the initial GDF map estimate}
Fixing the ribbon mask region for a given $(u,v)$, we then used the input map ENA rates $\hat M$ \textit{within and outside} the ribbon mask to predict the shape of the GDF \textit{under} the ribbon. Many different methods can be used to estimate a spatially-varying response surface \citep{Schwadron2018, Reisenfeld2021, Swaczyna2022}. We propose a multi-step approach for obtaining an initial GDF map estimate, where we first estimated a smooth spatial surface for the GDF using generalized additive modeling (GAM) \citep{Wood2017}. To fit the GAM, we used R package \textit{mgcv}, which allows users to specify the spatial model structure and degree of smoothing \citep{Wood2016}. We implemented an isotropic spherical analog to thin plate splines, which allows the fitted model to account for the spherical latitude/longitude structure of the data \citep{Wahba1981, Wood2003}. In order to incorporate information about the total ENA rate $\hat M$ within the ribbon mask region, we reformulated the problem in terms of interval censoring, where all GDF values in the mask region were assumed to be between 0 and $\hat M$, and fit the model using the interval-censoring compatible ``cnorm" link function. In order to enforce a large degree of smoothness on this fit, we incorporated first- to fourth-order derivative penalization and restricted to no more than 50 basis functions (15 for 6-degree maps). This GAM fit was implemented twice, where the second fit incorporated symmetry in the ENA rates above and below the axis of heliospheric motion. The initial smooth GDF prediction was a weighted average of the symmetric and asymmetric fits, where symmetry was up-weighted closer to the heliospheric tail. These symmetry assumptions were added in an attempt to improve discrimination between GDF and ribbon in the tail enhancement region (see \textbf{Supp. Figure G10}). \\
\indent The GAM model alone produces an estimate of the GDF macro-structure. In order to capture GDF micro-structure, we then considered the residuals between the input map $\hat M$ and the GAM model predictions outside the ribbon mask (or, for 2-degree input maps, a narrower ribbon mask). These residuals were then modeled using projection pursuit regression (PPR) \citep{Friedman1981} and the fitted residuals added to the GAM GDF map predictions. The GAM and PPR predictions together were then used as the initial estimate of the GDF map values inside the ribbon mask region. GDF map estimates were truncated to be between 0 and $\hat M$ as needed and were set equal to $\hat M$ outside the ribbon mask region. 

\subsubsection{Obtaining the fitted ribbon map estimate and re-estimating the GDF}
Subtracting the initial GDF map estimate from the input data $\hat M$, we obtained an initial estimate for the ribbon ENA rates within the mask region. However, this initial estimated ribbon map inherited any lack of fit or poor predictive performance of the GDF estimation, which sometimes resulted in various data artifacts and non-ribbon features evident in the ribbon. Ribbon cross-sections sometimes also changed shape radically across small spatial scales, which is not supported by physics. \\
\indent To address this issue, we estimated a new ribbon map by modeling the initial ribbon, imposing additional assumptions regarding the ribbon shape as illustrated in \textbf{Figure \ref{ribbonsepdiagram}}. Our ribbon modeling used a \textit{scaled} version of the initial ribbon map, where ENA rates in each azimuthal sector were divided the sector's estimated peak height. This scaling step allowed us to focus the modeling efforts on the ribbon shape/profile rather than its relative intensity, which also avoided unintentional and unrealistic smoothing of ribbon intensities across azimuthal sectors. The resulting fitted ribbon maps were much more robust to poor GDF map estimation and had fewer jagged artifacts. \\
\indent Our modeling of the ENA ribbon relied on two key assumptions: (1) each ribbon cross section is unimodal and (2) the \textit{shape} of the ribbon varies smoothly and not too quickly (i.e., on a spatial scale less than 15 degrees) across azimuthal angles. The former assumption relates to the ribbon shape (up to proportionality) within each cross-section, and the second assumption relates to the variability in this shape along the ribbon. We note, however, that (2) does not reflect any restrictions of the relative intensities of the ribbon across azimuthal sectors, which can and do change substantially on smaller spatial scales than 15 degrees. Briefly, we estimated the fitted ribbon map using a generalized additive model applied to the scaled initial ribbon map estimate as detailed in \textbf{Supp. Section G.2}, parameterized in terms of the azimuthal angle and the polar distance from the ribbon peak. After fitting, we then multiplied the fitted ribbon profile map by the azimuthal sector-specific maximum ENA rates to obtain the fitted ENA rate map. The resulting fitted ribbon map broadly satisfied the ribbon shape assumptions and had reduced sensitivity to noise and other data anomalies relative to the initial ribbon map. \\
\indent After ribbon map estimation, we then re-estimated the GDF by first subtracting the fitted ribbon from the input data map $\hat M$. GDF map estimation then proceeded broadly similarly to before (using GAM and PPR), but it did not mask out the ribbon region. The goal of this GDF map re-estimation was to ensure that the GDF had a fairly smooth structure in line with our physics understanding, while still allowing for some small scale GDF structure. This procedure resulted in an estimated GDF map, $\hat G$. Subtracting this estimate from the input data, $\hat M$, we obtained the ribbon map estimate, $\hat R$. 

\subsubsection{Estimating map uncertainties}
In order to estimate standard errors for the separated maps, we repeat the proposed algorithm many times, each time (1) drawing a realization of the input map in ecliptic coordinates assuming a multivariate normal distribution with mean $\hat D$ and covariance $\hat \Sigma_D$, (2) rotating the realization into ribbon-centric coordinates, (3) generating a bootstrap sample of this realization stratified by azimuthal angle (i.e., by longitude bin), and (4) performing the proposed ribbon separation method. The set of resulting separations will incorporate variability both due to the input map uncertainty (from the multivariate normal draw) and due to the separation uncertainty for a fixed input map (from the bootstrapping). The pixel-wise standard deviation and pixel cross-correlations calculated across the replicates provide Monte Carlo-type uncertainties for the separated maps. We can view the total uncertainty of a separated/partitioned map as resulting (a) \textit{aleatoric/statistical} uncertainty of the input map and (b) \textit{epistemic/systematic} uncertainty due to ambiguity in the ribbon/GDF partitioning. We account for these two sources of uncertainty jointly using multivariate normal draws and bootstrap sampling, respectively.\\
\indent Ideally, the proposed uncertainty estimation algorithm would be repeated for many draws of the input maps. Due to the computational burden associated with performing so many separations across tuning parameter specifications and drawn/bootstrapped input maps, however, a very large number of replicates is computationally infeasible. In practice, we calculated separation uncertainties using 250 replicates of the above algorithm, and a more detailed exploration of these results is provided in \textbf{Supp. Section B}.

\section{Evaluating separation and center estimation in simulated data} \label{simresults}
We evaluated the performance of the proposed map separation and ribbon center estimation methods in simulated data, under three different simulation scenarios:
\begin{enumerate}[noitemsep, label=S{\arabic*}.]
    \item Weak scattering model 
    \item Spatial retention model 
    \item Smoothly varying ribbon profile between spatial retention and weak scattering,
\end{enumerate}
where physics simulation models were used to generate the spatial retention and weak scattering (i.e., the two leading ribbon formation hypotheses) profile shapes. In each scenario, the true total ENA rate map was defined as the sum of separately-simulated globally-distributed flux and ribbon maps. For the first two simulation scenarios, the globally-distributed flux map was generated using a scaled version of the 2.73 keV GDF simulation from \citet{Zirnstein2017}. Corresponding ribbon-only maps were generated to mimic the expected shape of the ribbon under two leading ribbon generation hypotheses as detailed in \citet{Zirnstein2019b}, where the ribbon shape was determined using a single representative simulation cross-section and where the ribbon was assumed to be exactly circular and centered at longitude 221.5 and latitude 39. Dimming was added to certain sections of the ribbon to mirror observed IBEX data. For scenario S3, we considered a more complicated structure for the ribbon and GDF maps, including a spatially-varying ribbon shape, spatially concentrated knots on the ribbon, and a large high-flux disc on the GDF. Scenario S3 is not believed to be realistic; rather, this simulated scenario was devised to help characterize the limitations of the proposed methods. \\
\indent Under each of scenarios S1-S3, we performed ribbon center estimation and separation using several different types of pixelated total ENA rate maps. First, we considered ``ideal" input maps equal to the sum of the simulation truth ribbon-only and GDF-only maps. Unlike these idealized input maps, any ENA rate maps generated using \textit{real} IBEX data must overcome additional challenges including substantial blurring due to the IBEX-Hi instrument field of view, undesired background, and missing data. To evaluate our methods using simulated ENA maps constructed from more realistic data settings, we simulated ENA data mimicking IBEX observed data, including field of view-related spatial blurring and known levels background interference. Five simulated data replicates were generated, each using exposure times, observation locations, and background rates observed in the 2018 IBEX data for energy steps (ESAs) 2-6. Using these messy observed data, we applied the Theseus map construction strategy to obtain pixel-level ENA rate estimates and uncertainties for each energy step and input map combination \citep{Osthus2022}. Results based on these messier ``estimated" input maps inherit any errors or biases induced by the map-making procedure. Finally, we also estimated maps based on simulated data with 3 times longer observation time for each pixel, resulting in a third set of ``estimated, 3x" input maps. The motivation for including this third set of maps was to mimic an emerging IBEX data product expected to triple the number of observed ENA encounters used for map-making (unpublished).

\subsection{Ribbon center estimation performance}
\begin{figure}[h!]
  \centering
\subfloat[Bias of ribbon center estimates (degrees; lower = better)]{\includegraphics[trim={0cm 0cm 0cm 0.6cm}, clip, width=6in]{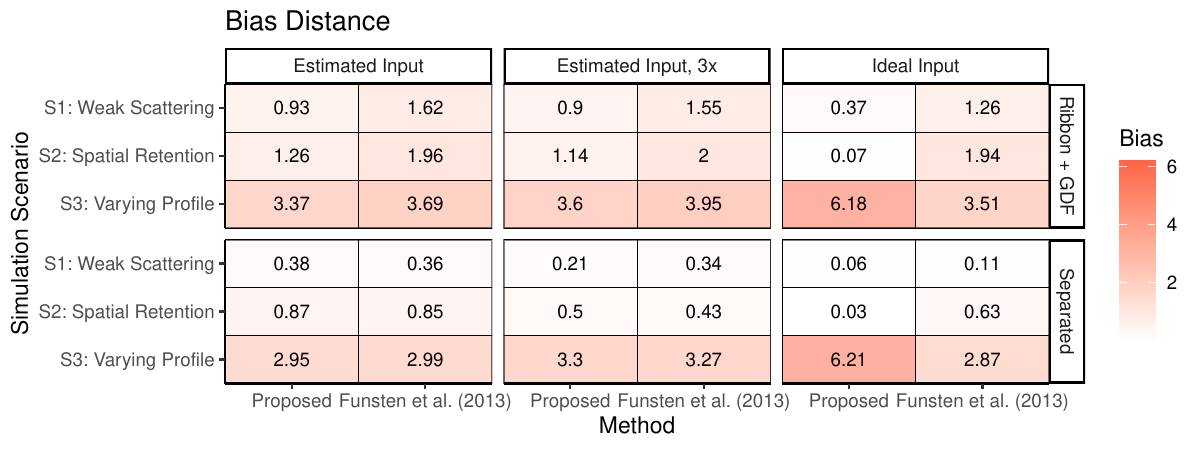}}\\
\hspace{0.8in}\subfloat[Bias as a function of working rotational frame for true ribbon-only spatial retention map (scenario S2)]{\includegraphics[trim={0cm 0cm 0cm 0.8cm}, clip, width=5.2in]{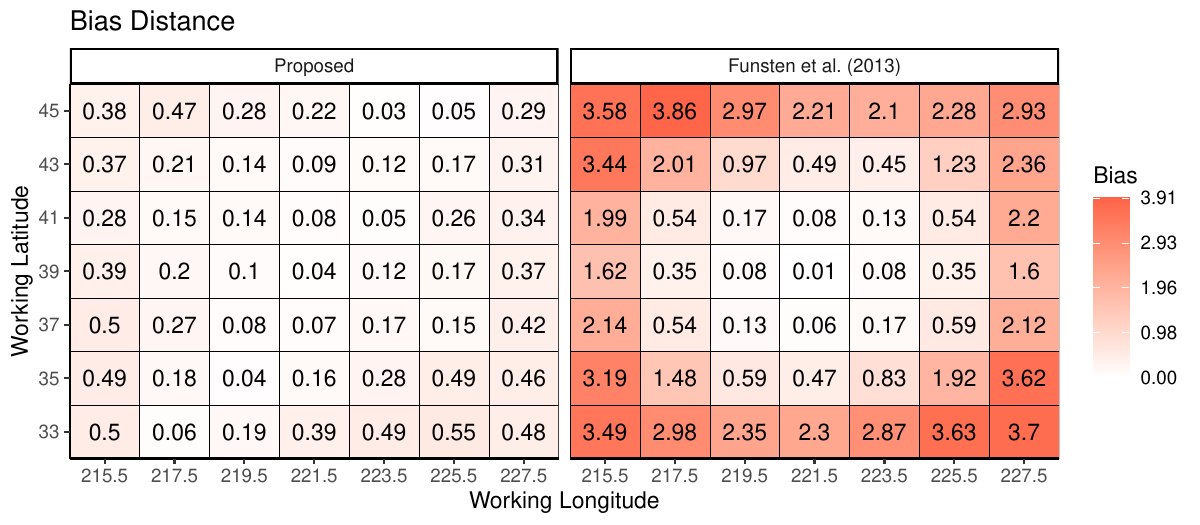}}
\caption{Performance of ribbon center estimation in simulated data\\
 \footnotesize This first panel of this figure shows the Euclidean distance (i.e., bias) between ribbon center estimates and (221.5, 39), for ribbon center estimates obtained using the proposed method in \textbf{Section \ref{center}} and using the method in \citet{Funsten2013}. Lower values indicate more accurate ribbon center point estimates. All estimates are calculated using the correct rotational frame, where the true ribbon center in each simulated map is located at (221.5, 39) in ecliptic coordinates. In this panel, ``Ribbon + GDF" indicates estimation on the estimated or simulation truth map with both ribbon and GDF and ``Separated" refers to center estimation using maps that have been ribbon separated using the method in \textbf{Section \ref{separation}}. The second panel shows the ribbon center bias after 1 iteration of each center estimation algorithm under different working values for the ribbon, for a true simulated ribbon-only spatial retention map. }
 \label{centerest}
\end{figure}

\indent For each set of maps under each of the three ribbon profile scenarios, we applied the proposed ribbon center method described in \textbf{Section \ref{center}} to estimate the location of the ribbon center and its uncertainty. For comparison, we also implemented the method proposed in \citet{Funsten2013}, which has several drawbacks relative to our proposed method. In particular, it does not account for input map estimation uncertainty, it models the ribbon profile using a Gaussian structure, and it performs the ellipse estimation in the working ribbon center frame only. Unless otherwise specified, the working ribbon center was set equal to the simulation truth. In all scenarios, the \textit{true} ribbon center was located at longitude 221.5 and latitude 39 in the ecliptic frame. \\
\indent In nearly all scenarios, the proposed center estimation method resulted in similar or lower bias and more reasonable uncertainty estimates than the method from \citet{Funsten2013}. \textbf{Figure \ref{centerest}a} shows the ribbon center bias (degrees) when estimation used the \textit{correct} working frame. Corresponding uncertainty ellipses are shown in \textbf{Supp. Figure D1}. For scenarios S1/S2, error in ribbon center estimates were less than 0.5 degrees for all ``ideal" (i.e., not estimated) simulated maps. Ribbon center estimation using the total flux map (i.e., with Ribbon + GDF) resulted in slightly higher estimation error than ribbon center estimation using separated ribbon-only maps. Uncertainty ellipses for the proposed method tended to be substantially larger than those estimated by \citet{Funsten2013}, since the proposed methods captured uncertainty due to both the input map estimation and the ribbon center estimation for a fixed input map.\\
\indent With ideal input maps, the method in \citet{Funsten2013} outperformed the proposed method for scenario S3. The proposed method produced more accurate estimates of the ribbon peaks; however, the ribbon peaks turned out to be a poor proxy for ribbon location in scenario S3, with better peak estimates counter-intuitively leading to worse center estimates. This result provides a cautionary example highlighting that different metrics may be needed for characterizing the ribbon location when the ribbon changes shape/skewness dramatically across azimuthal angles. This problem is explored in \textbf{Supp. Section D}. We demonstrate that an alternative ribbon location metric based on the full width at quarter max (i.e., the ribbon width at 25\% of the cross-section peak) may outperform peak-based center estimates in some settings.  \\
\indent All estimates in \textbf{Figure \ref{centerest}a} assumed that ribbon center estimation was performed using correct working rotational frame, but we would not expect this to be correctly specified in general. When the working center was poorly specified (e.g., by 4 degrees in any direction), the \citet{Funsten2013} center estimates showed substantial bias (\textbf{Figure \ref{centerest}b}); this bias was not seen under the proposed method because it estimates ribbon centers in the best-fitting plane rather than the working rotational frame. Sensitivity to the working rotational frame was substantially mitigated by iterating the ribbon center estimation (\textbf{Supp. Figure D4}). Weighted interval scores for estimates in \textbf{Figure \ref{centerest}} (calculated similarly to \textbf{Section \ref{acc}}) are provided in \textbf{Supp. Figure D2}. 

\subsection{Ribbon separation performance}
\indent For each simulation scenario and each simulated input map, we applied the proposed map partitioning method described in \textbf{Section \ref{separation}}. For comparison, we also implemented the ribbon separation methods proposed in \citet{Reisenfeld2021}. Both algorithms performed ribbon separation in the correct working rotational frame. For the proposed ribbon separation approach, separation was repeated across 63 combinations of mask tuning parameters, with the final separation obtained as a weighted combination of the best 25\% of separations according to our goodness of separation heuristic. We used a multi-faceted approach for assessing ribbon separation performance, including visual inspection of separated maps (\textbf{Section \ref{vis1}}); percent error, correlation, and weighted interval score \citep{Bracher2021} comparing separated GDF map estimates and simulation truths (\textbf{Section \ref{acc}}); ribbon skewness for separated ribbon maps and simulation truths; and visual comparison of estimated ribbon morphology (\textbf{Section \ref{skewsims}}). Additional evaluations are provided in \textbf{Supp. Section E}. Bias, correlation, WIS, and coverage diagnostics are provided using maps in the ribbon-centric rotational frame, and complementary diagnostics calculated in the ecliptic rotational frame are provided in \textbf{Supp. Figure E8}. 

\subsubsection{Visual evaluation of ribbon separated maps}\label{vis1}
\begin{figure}[h!]
  \centering
\includegraphics[trim={0cm 0cm 0cm 0.6cm}, clip, width=6in]{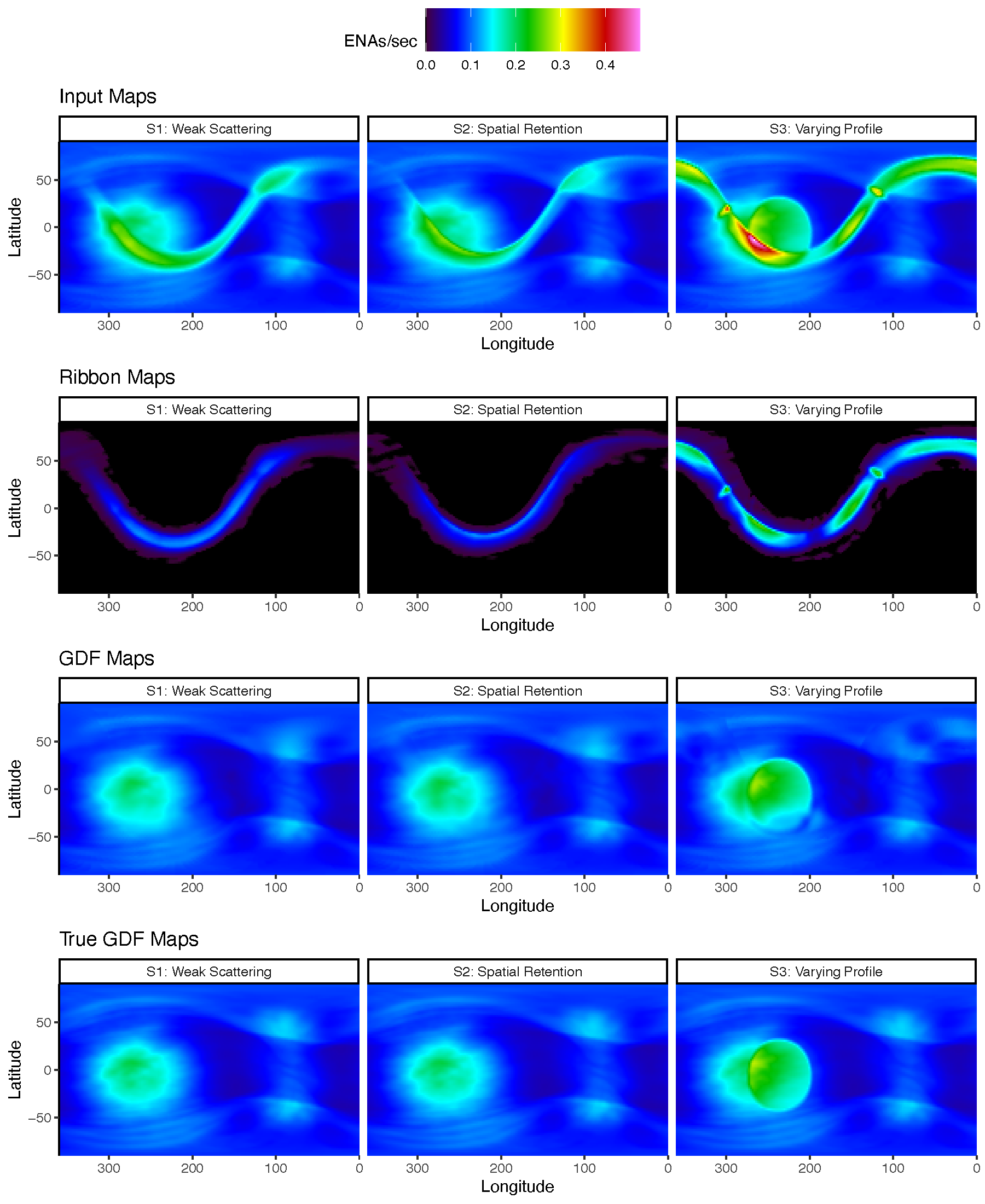}
\caption{Estimated ribbon and GDF maps obtained using the proposed ribbon separation method applied to ``ideal" (not estimated) simulated input maps\\
\footnotesize This figure shows simulated map partitions obtained using the method proposed in \textbf{Section \ref{methods}}. Results are shown in the ecliptic rotational frame. For comparison, the true GDF-only maps are shown in the final row of the figure.}
\label{simsep_ecliptic}
\end{figure}

\indent In \textbf{Figure \ref{simsep_ecliptic}}, we present the GDF and ribbon-only maps obtained by applying the proposed methods to each of the three ``ideal" input datasets. Corresponding maps in the ribbon-centric frame and/or based on estimated input maps are presented in \textbf{Supp. Figure E1}. In scenarios S1/S2, the ribbon separations appear very clean, with little evidence of ribbon-related artifacts in the GDF-only map. For scenario S3, the region formerly containing the ribbon is visible in the GDF maps through very minor residual blurring and other artifacts. As shown in \textbf{Supp. Figure E2}, maps produced using the methods in \citet{Reisenfeld2021} show much more pronounced banding and other artifacts. 

\subsubsection{Accuracy and uncertainty of GDF maps} \label{acc}
\begin{figure}[h!]
\subfloat[Average absolute percent error between true and estimated GDF maps (lower = better)]{\includegraphics[trim={0cm 0.5cm 0cm 0.8cm}, clip, width=6in]{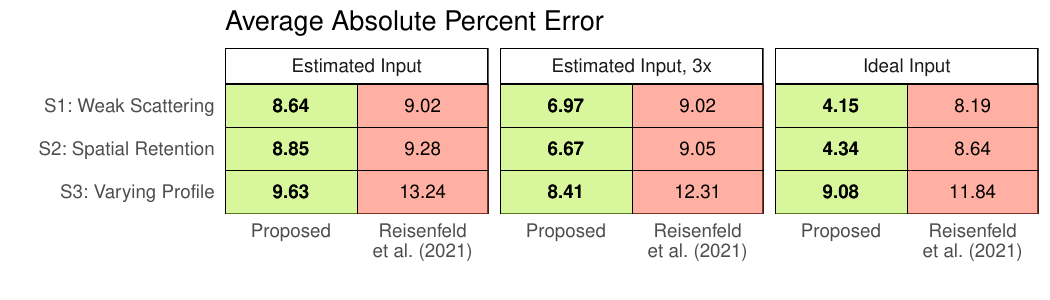}}\\
\subfloat[Spearman correlation between true and estimated GDF maps (higher = better)]{\includegraphics[trim={0cm 0.5cm 0cm 0.7cm}, clip, width=6in]{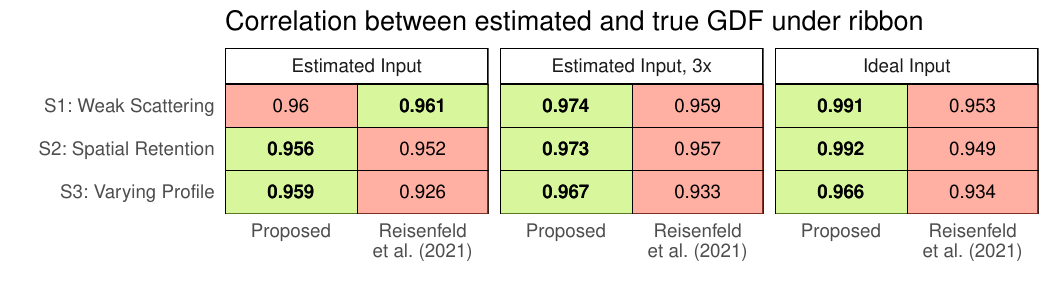}}\\
\subfloat[Average weighted interval score between true and estimated GDF maps (lower = better)]{\includegraphics[trim={0cm 0.5cm 0cm 0.8cm}, clip, width=6in]{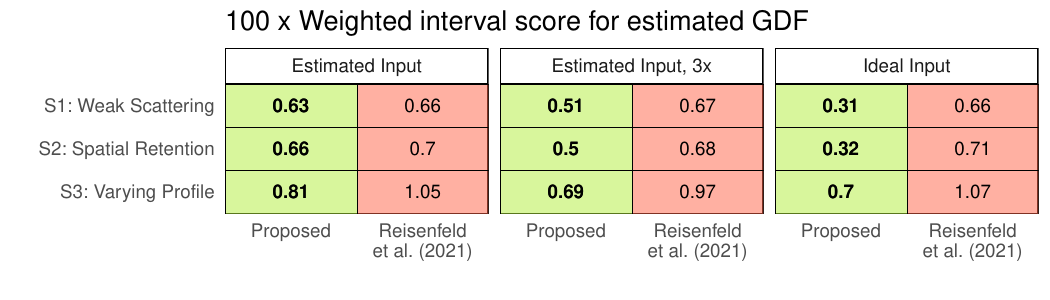}}\\
\subfloat[Coverage of 95\% confidence intervals of GDF map estimates (closer to 0.95 = better)]{\includegraphics[trim={0cm 0.5cm 0cm 0.8cm}, clip, width=6in]{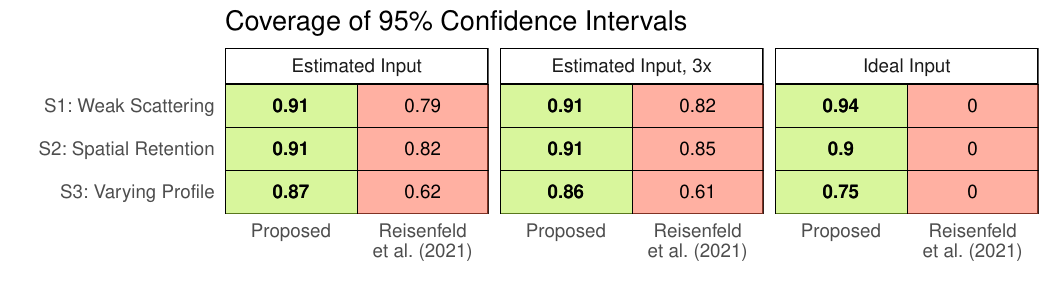}}
\caption{Performance of ribbon separation methods in terms of comparisons between estimated and true $GDF$ maps in the ribbon region}
\label{simerror}
\end{figure}

\indent We provide a more rigorous evaluation of the simulated map separation in \textbf{Figure \ref{simerror}}. Overall, the proposed ribbon separation performed well, outperforming the existing method in \citet{Reisenfeld2021} across nearly all metrics and scenarios considered.\\
\indent In the first panel of \textbf{Figure \ref{simerror}}, we present the average percent error in GDF estimates in the ribbon region, and we provide the correlation between the true and estimated GDF values in the second panel. The proposed method resulted in high correlation between true and estimated GDF values across all simulation scenarios. Correlations for the method in \citet{Reisenfeld2021} were also high, but appreciably lower than the proposed method for most maps. The proposed method also resulted in much lower absolute percent error (by 25-50\% for ideal inputs) than the method in \citet{Reisenfeld2021} across all simulation scenarios and input datasets considered. \\
\indent The first two panels of \textbf{Figure \ref{simerror}} focus on estimated GDF and ribbon maps, but we are also interested in evaluating their standard errors. In the third panel, we present the estimated GDF map weighted interval score, which is a function of both the estimation error and corresponding uncertainties. Lower values of this score indicate better performance. We calculated the weighted interval score for pixel $i$ using the equation:
\begin{align}\label{wisGDF}
\text{WIS}_i  &= \frac{1}{M+0.5}\left[\frac{1}{2} \vert \hat{G}_i - \mathcal{G}_i \vert + \sum_{m=1}^M \frac{\alpha_m}{2} \vert \hat u_{im} - \hat l_{im} \vert  + \sum_{m=1}^M (\hat l_{im}-\mathcal{G}_i )^{ I(\mathcal{G}_i<\hat l_{im})} (\mathcal{G}_i-\hat u_{im} )^{I(\mathcal{G}_i>\hat u_{im})} \right] \nonumber
\end{align}
where $\hat G_i$ corresponds to the estimated GDF map, where $\mathcal{G}_i$ is the true value of the GDF, and where $\hat l_{im}$ and $\hat u_{im}$ are the $(1-\alpha_{m}/2)$\% lower and upper confidence interval estimates, respectively. This weighted interval was averaged for all pixels with polar angles between 80 and 130, the region where both ribbon and GDF are potentially present. The weighted interval scores were calculated using $\alpha = (0.02, 0.05, 0.1, 0.2, ..., 0.8, 0.9)$.  The proposed method produced lower values for all simulation scenarios and input map formulations. \\
\indent The fourth panel of \textbf{Figure \ref{simerror}} provides coverage of 95\% confidence intervals for the GDF map estimates across pixels and 5 estimated map replicates. Ideal input map coverages for the \citet{Reisenfeld2021} are zero as that method propagates only uncertainty in the input map. Coverages for all scenarios (and scenario S3 in particular) were below the nominal 0.95. Under-coverage was a combination of two main factors: (1) error in the ribbon separation and (2) slight under-coverage of the estimated input map uncertainties. This issue is discussed in \textbf{Supp. Section B}. In spite of this, our results demonstrate similar or better statistical coverage relative to the method in \citet{Reisenfeld2021}. 


\subsubsection{Visual comparison of ribbon profile shapes and skewness diagnostics} \label{skewsims}
\begin{figure}[h!]
  \centering
\subfloat[Ribbon Map Estimates]{\includegraphics[trim={0cm 0cm 0cm 0cm}, clip, width=6in]{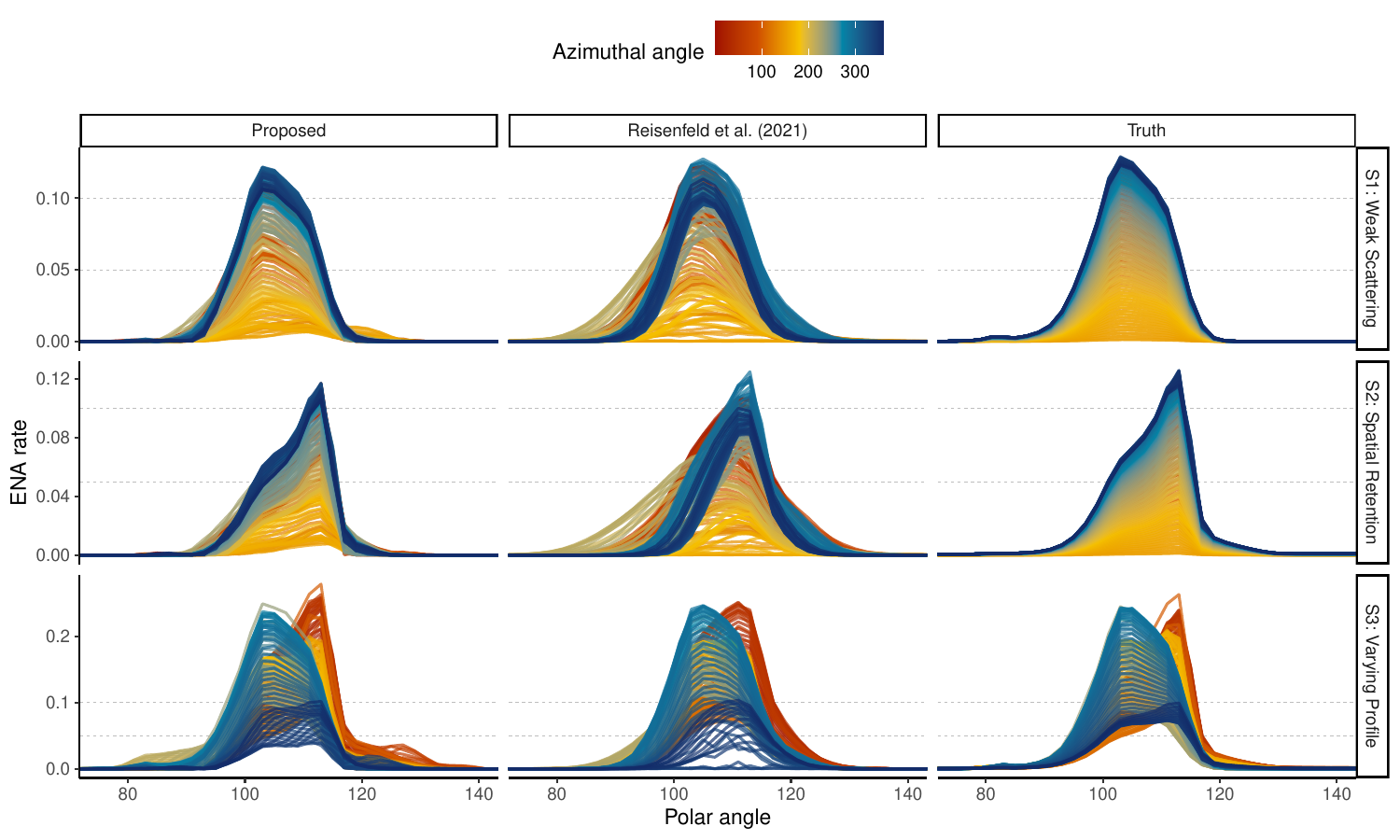}}\\
\subfloat[GDF Map Estimates]{\includegraphics[trim={0cm 0cm 0cm 2cm}, clip, width=6in]{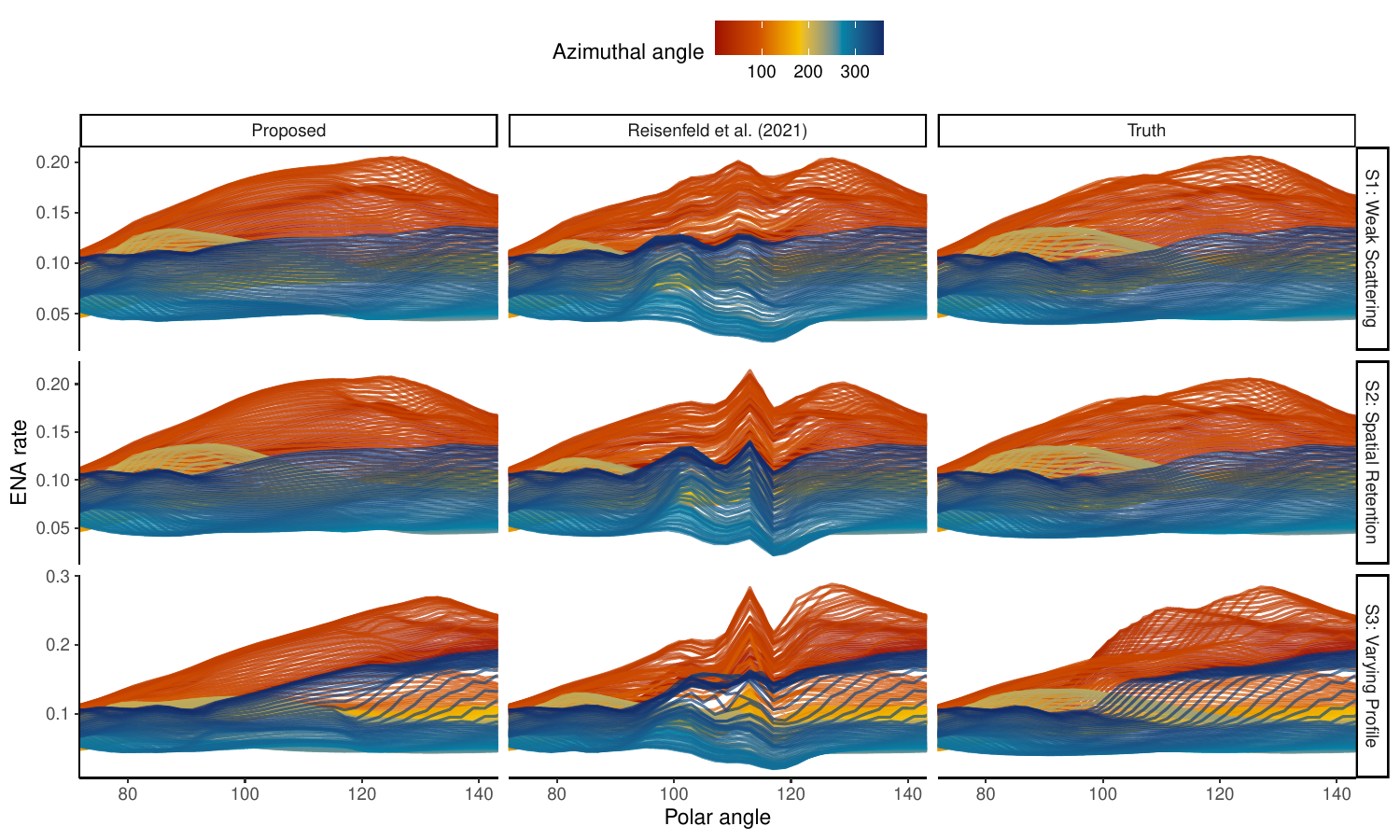}}
\caption{Spaghetti plots of ribbon/GDF map cross-sections by azimuthal sector for ideal (not estimated) input maps across different simulation scenarios. Results are shown using the proposed method and the method in \citet{Reisenfeld2021}. \\
\footnotesize This figure shows the estimated $\hat{\hat R}_i$ and $\hat{\hat G}_i$ values obtained for each simulation scenario, using ideal (i.e., not estimated) input maps. Lines correspond to cross-sections of ENA rates across different azimuthal sectors within each map.  }
\label{simprofilesest}
\end{figure}

\indent In addition to accuracy of the individual pixels in the separated maps, we are also interested in the profile or shape of the estimated ribbon. This is particularly relevant to hypothesis testing between competing physics models explaining ribbon generation. \\
\indent \textbf{Figure \ref{simprofilesest}} compares ribbon cross-section estimates obtained using the proposed methods to simulation truth for each azimuthal sector. Using ideal input maps, the proposed separation method captures the bent ``shoulder" structure for scenario S2 and results in high fidelity to simulation truths for scenarios S1/S2. In scenario S3, the ribbon cross-sections do a good job of capturing the changing shape of the ribbon but do a poor job at estimating the shape for azimuthal angles above 300. The fitted ribbon shapes from the proposed method are much more accurate than those obtained using the method in \citet{Reisenfeld2021}, which tends to smooth over nuance. In all simulation scenarios, ribbon profiles based on estimated (rather than ideal) input maps did not clearly have these distinguishing features when not present in the input map. \\
\indent Skewness could be potentially useful for discriminating between competing ribbon generation theories, which are associated with different ribbon skewness properties \citep{Zirnstein2019}. \textbf{Supp. Figure E5} compares estimated skewness in our separated ribbon maps with the simulation truth. \citet{Reisenfeld2021} produced near-symmetrical ribbon maps, making hypothesis discrimination based on skewness futile for those maps. The proposed maps resulted in skewness estimates that were biased toward zero relative to the simulation truth but that did capture differences between the two physics models in terms of skewness for ideal input maps. Skewness-based discrimination between physics model was less clear for estimated input maps. Similar diagnostics for the ribbon widths (full width at half max) are provided in \textbf{Supp. Figure E6}.

\section{Application to real IBEX data} \label{mainrealdata}
\begin{figure}[h!]
  \vspace{-0.5cm}
  \centering
\subfloat[Input Map]{\includegraphics[trim={3.3cm 3.1cm 2.7cm 2.55cm}, clip, width=2in]{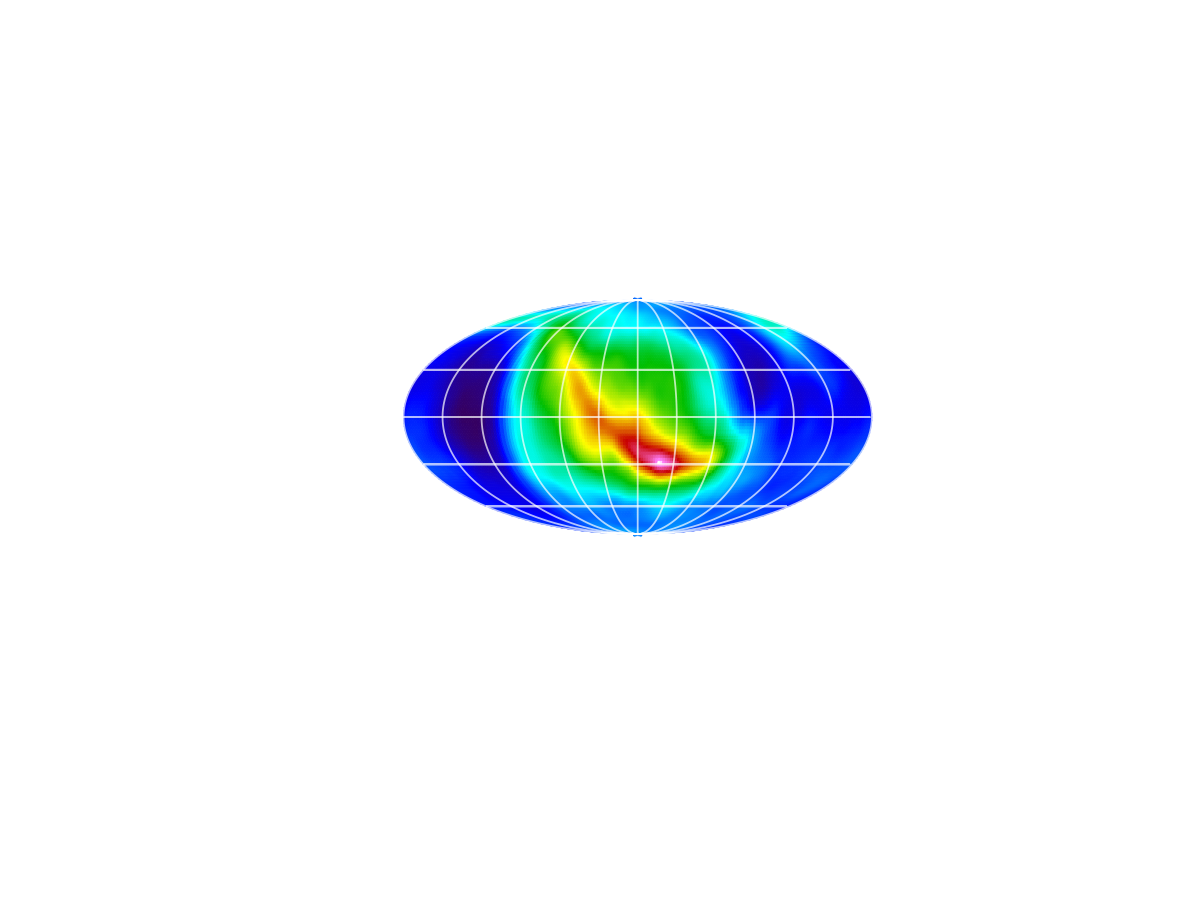}}
\subfloat[Ribbon Map]{\includegraphics[trim={3.3cm 3.1cm 2.7cm 2.55cm}, clip, width=2in]{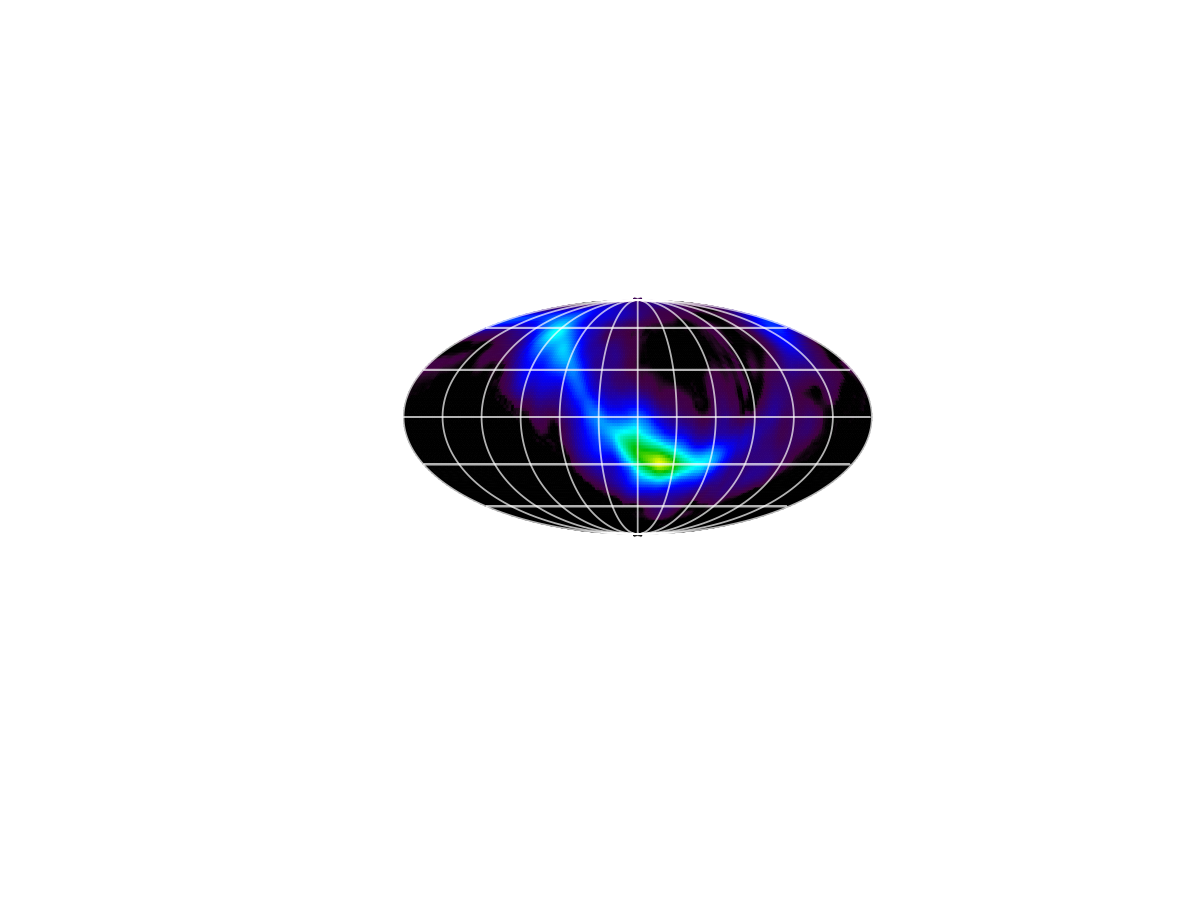}}
\subfloat[GDF Map]{\includegraphics[trim={3.3cm 3.1cm 2.7cm 2.55cm}, clip, width=2in]{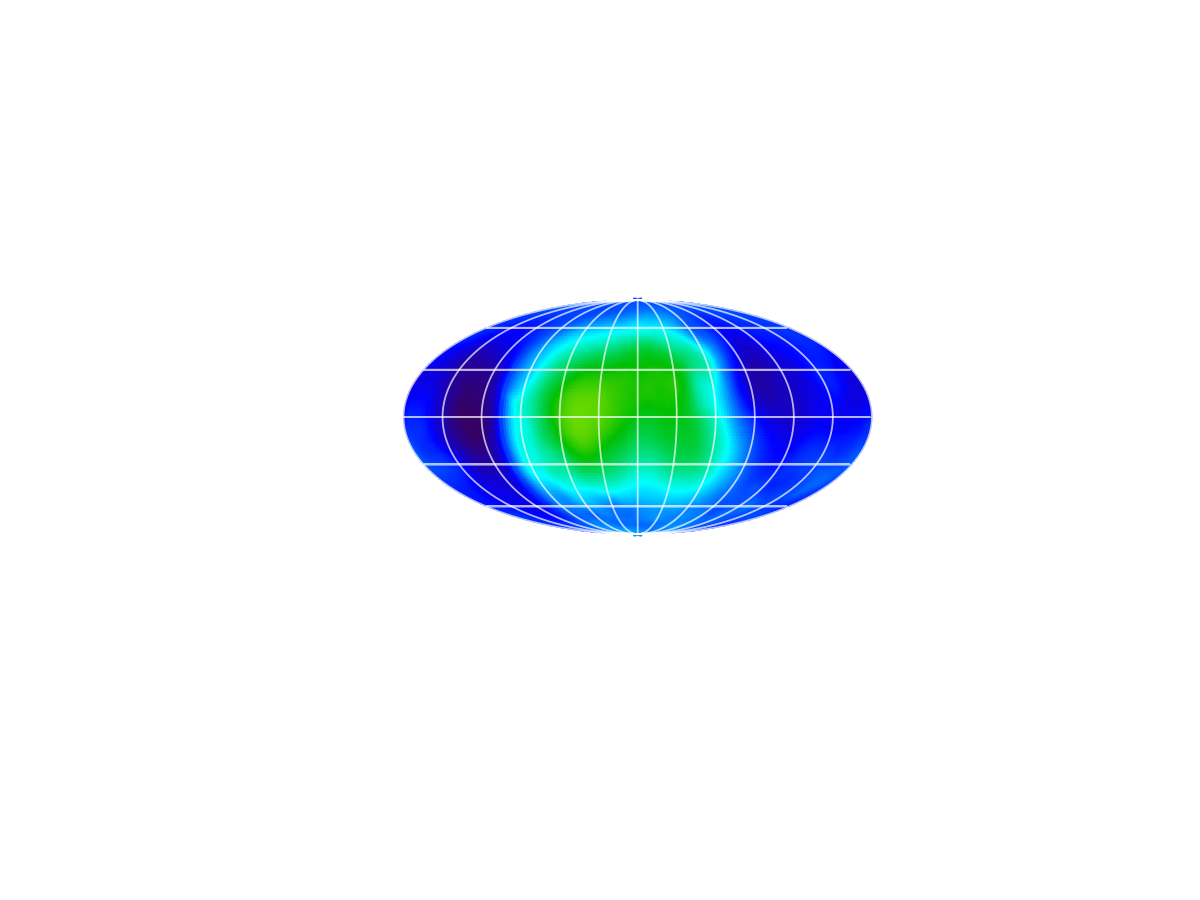}}\\
\includegraphics[trim={0cm 2cm 0cm 1.5cm}, clip, width=3in]{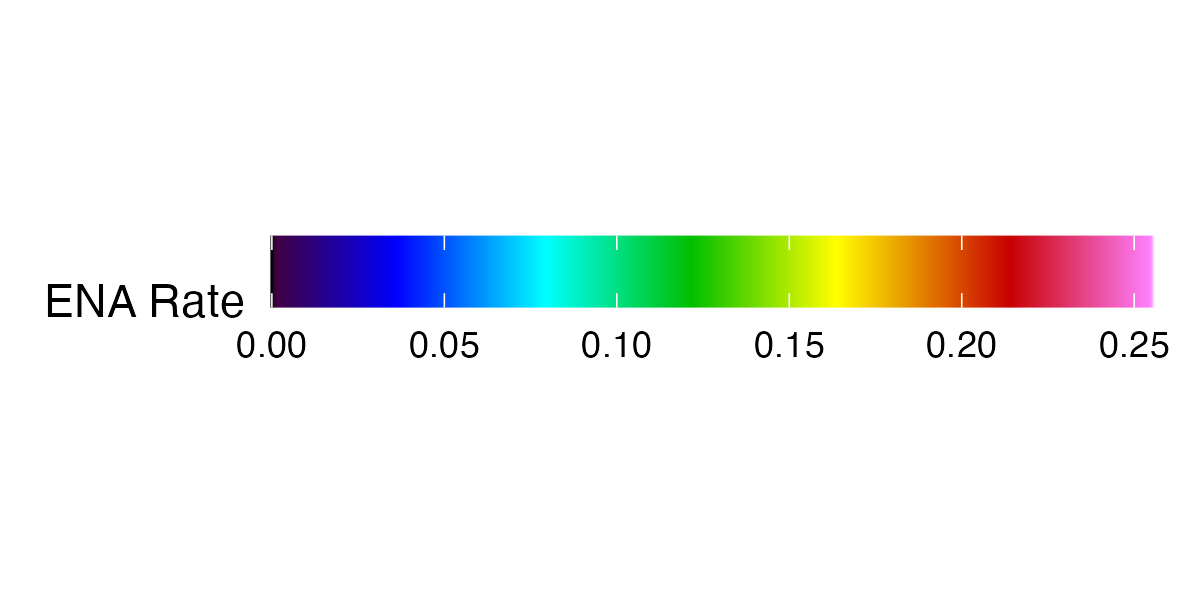} \hspace{1cm}
\caption{Example separated maps for real IBEX data Theseus map\\
 \footnotesize This figure shows the results of ribbon separation for 2-degree Theseus input maps generated using real IBEX data for the first half of 2019, energy step 4. Results are based on maps in the spacecraft frame do not include ENA capture efficiency corrections or Compton-Getting effect corrections \citep{Compton1935}. The large enhancement in ENA rates in (c) corresponds to the ``nose" of the heliosphere (i.e., the direction in which the Sun and heliosphere is moving with respect to interstellar space.  }
 \label{ex_realdata}
  \vspace{-0.5cm}
\end{figure}

\begin{figure}[h!]
  \centering
\subfloat[Skewness and width (full width at half max) of estimated ribbon profiles by azimuthal sector across 60 separated ram maps (10 maps per energy step)]{\includegraphics[trim={0cm 0cm 0cm 0.1cm}, clip, width=6in]{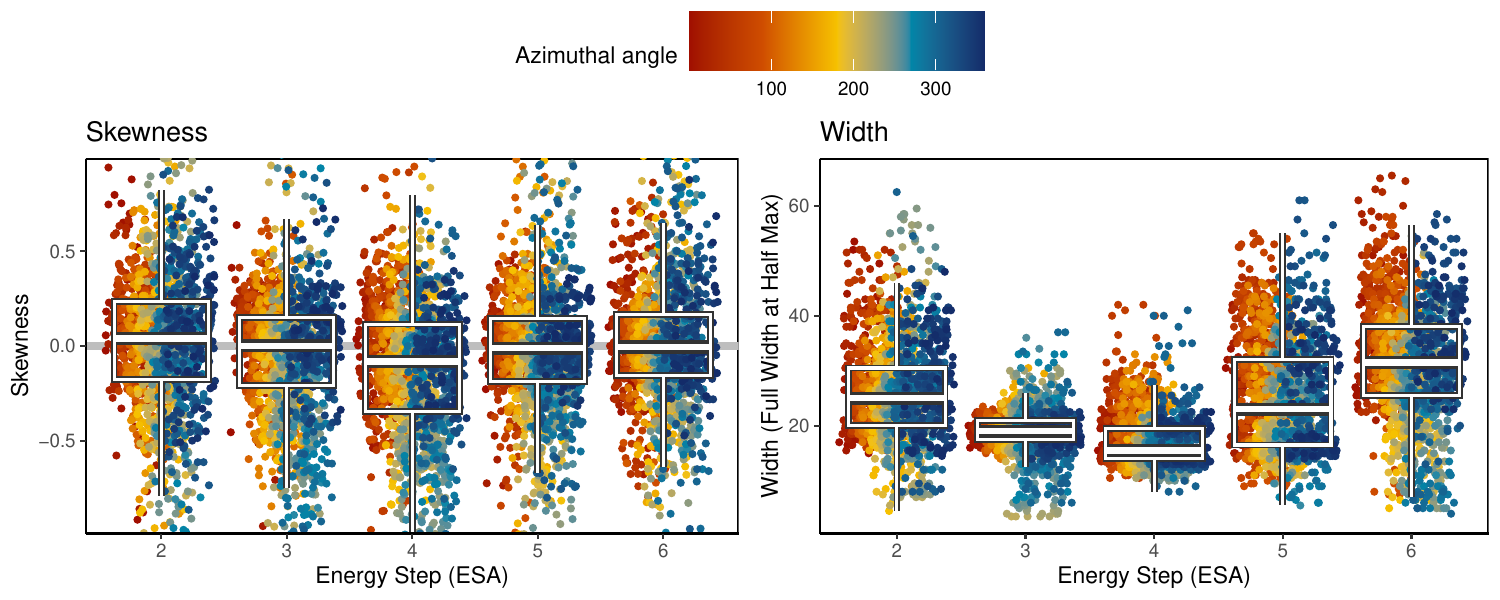}}\\
\subfloat[Maximum estimated ribbon ENA rates over time by energy step (ESA)]{\includegraphics[trim={0cm 0cm 0cm 0.8cm}, clip, width=6in]{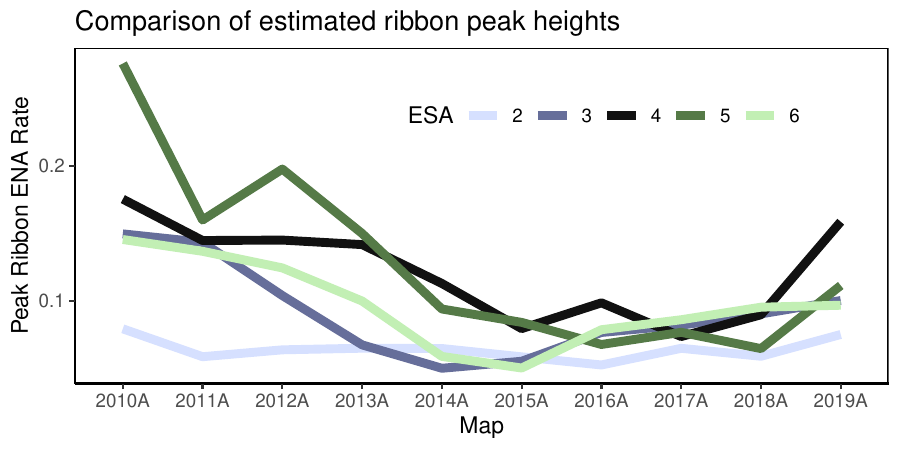}}\\
\caption{Estimated ribbon shapes using real IBEX data by energy step (ESA) \\
 \footnotesize The first panel shows the estimated skewness and width (full width at half max) of ribbon map profiles for each of 180 azimuthal sectors for each of 60 2-degree maps generated using real IBEX data (first 6-months of each year between 2010 and 2019, energy steps 2-6). The full width at half max was estimated by fitting a cubic spline interpolation for each azimuthal angle and identifying the most extreme polar angles (at 0.01 degree polar angle bins) with predictions at least 50\% of the azimuthal sector max. The second panel shows the point estimate for the highest estimated ribbon ENA rates for each map, where 'A' indicates a map using data from the first 6 months of each year.  }
 \label{skewness_realdata}
\end{figure}

\indent We applied the proposed ribbon center estimation and separation methods to 2-degree real data ``Theseus" maps generated using the method in \citet{Osthus2022} and to 6-degree real data maps generated by the IBEX Science Operations Center (ISOC). In this work, we provide some simple visual illustrations of the results of ENA ribbon separation applied to ENA rate maps. Ribbon center estimates along with detailed explorations of the ribbon-only and GDF-only maps will be presented in a series of follow-up papers.  We apply the ribbon separation to ENA rate maps generated using data collected between 2010 and 2019 and across 5 energy step bins (called ESAs 2-6). ESA 2 corresponds to lower energy $\sim$0.7 keV ENAs, while ESA 6 corresponds to higher energy $\sim$4.3keV ENAs.\\
\indent  \textbf{Figure \ref{ex_realdata}} shows the ribbon separation using the proposed method for data measured in the first half of 2019 for energy step 4. This map is a representative example of the map separation performance for energy steps 3-5. All separated maps for energy step 4 are provided in \textbf{Supp. Figures F2-F3}, and separations for other energy steps are provided in a GitHub repository at \url{https://github.com/lanl/IBEX_RibbonSeparation} and for 2019 in \textbf{Supp. Figures F4-F5}. Visual assessment reveals minimal banding or other artifacts in the ribbon region. Crucially, the separated maps appear to preserve some fine-scale features of the underlying GDF. Ribbon separation was most challenging for energy steps 2 and 6. For energy step 6 in particular, the estimated ribbon maps were somewhat patchy, indicating poor discrimination between GDF and ribbon for these maps (not shown). Data collected at this energy step do not have a defined ribbon to separate, and poor discrimination for energy step 6 was also seen using the method in \citet{Reisenfeld2021} (not shown).\\
\indent \textbf{Figure \ref{skewness_realdata}a} provides the ribbon skewness and width (measured by full width at half max) for the ribbon-only map estimates within each azimuthal angle across all Theseus maps considered. These results reveal a fairly symmetrical ribbon on average, with possibly some slight negative skewness (more ENA flux toward ribbon center) for energy steps 3-5. Estimated widths were the smallest for step 4 on average, with median estimates of 25.0, 19.0, 15.5, 23.0, and 31.5 degrees, respectively, for energy steps 2-6. Similar results are provided for ISOC maps in \textbf{Supp. Figure F6}, with estimated median ribbon widths between 20 and 27 degrees. The near-zero ribbon skewness observed for these maps provides some evidence in favor of symmetric ribbon profile theories such as the weak scattering hypothesis. Additional analysis using corrected versions of these data are warranted to evaluate this question more rigorously.\\
\indent \textbf{Figure \ref{skewness_realdata}b} provides the maximum estimated ribbon ENA rate over time by energy step. The method in \citet{Reisenfeld2021} tended to produce slightly dimmer ribbons than the proposed method, likely due to the more restrictive ribbon shape assumed by \citet{Reisenfeld2021}. However, changes in the ribbon heights over time and by energy tended to be qualitatively similar between the methods. Both methods indicated that the ribbon intensity changed dramatically between 2009 and 2019, with much weaker ribbon flux from roughly 2015 to 2018. The increased ENA rates observed at the start of 2019 are consistent with the hypothesis that the peak ribbon ENA rate follows the 11 year solar cycle, with a delay due to secondary ENA dynamics in the outer heliosphere and travel time of ENAs back to Earth \citep{McComas2020}.

\section{Discussion} \label{discuss}
Humanity is just beginning to understand the nature of the heliosphere and the heliospheric boundary between our solar system and interstellar space. Energetic neutral atoms (ENAs) may provide insight into the temporal and spatial variation in the shape and location of the boundary heliosheath region. Through the Interstellar Boundary Explorer (IBEX) mission, NASA collects data on the intensity of these ENA particles over the full sky. These data can be used to construct pixelated maps of the observed ENA intensities as a function of calendar time and binned ENA energy. \\
\indent ENAs captured by these maps are believed to primarily originate from two separate processes, and these two sources of ENAs and corresponding ENA intensities are each of scientific interest. Therefore, a key step of IBEX data processing is partitioning into separate maps representing both ENA types: globally-distributed flux (GDF) and the ENA ``ribbon." In this work, we proposed a new statistical algorithm for partitioning ENA maps into source-specific maps. Ancillary to this goal is estimation of the so-called ribbon ``center", and we also developed methods to perform this estimation. The performance of the proposed methods was evaluated using simulated data designed to mimic observed IBEX data and compared against leading alternative methods in the scientific literature. Selected real and simulated data results presented in this paper can be reproduced using code and data available in a companion Github repository at \url{https://github.com/lanl/IBEX_RibbonSeparation}. \\
\indent We found that the proposed ribbon separation algorithm had excellent performance in terms of the accuracy of map partitioning and the ability to capture nuanced ribbon shape. The gains over the ribbon separation method in \citet{Reisenfeld2021} substantially increased as the quality of the input data improved, primarily because key subtle features of ribbon morphology where smoothed out or entirely absent in lower-quality maps. In terms of IBEX mission data processing, our results support development and use of high-quality, high-resolution IBEX data products such as the finely resolved 2-degree ENA maps developed in \citet{Osthus2022}. One limitation of the proposed method (and likely any ribbon separation method) is its tendency to produce slightly over-dispersed ribbon profiles by construction (\textbf{Figure \ref{simprofilesest}} and \textbf{Supp. Figure E6}). Future evaluation of dispersion metrics should keep this limitation in mind.  \\
\indent In addition to map separation, we also proposed new methods for estimating the ``center" of the ENA ribbon and these methods showed good performance in simulated data analyses. Our results demonstrated that iteration is key; a single iteration of ribbon center estimation can result in substantial residual bias for both the proposed method and, to an even greater degree, the method in \citet{Funsten2013}. Our results also suggest that care should be taken in how the ``location" of the ribbon is defined in three-dimensional space, because the ribbon peak location can produce misleading results when the skewness of the ribbon varies dramatically for different ribbon regions. \\
\indent Our initial application of these methods to real IBEX data indicated that ENA ribbon cross-sections tended to have a small or slightly negative skewness for energy passbands (ESAs) 3-5. The width of the ENA ribbon also varied by ESA, with narrowest ribbon profiles observed for ESA 4 on average. Temporal changes in the ribbon intensity were also observed, mirroring the cadence of the solar cycle. Future work will explore implications of these temporal and energy passband-related variations in more detail. We emphasize that Theseus map real data results presented herein are for illustration and should not be used for subsequent heliospheric science directly, because additional data corrections and de-biasing steps are necessary before final estimates of ribbon and GDF properties can be obtained. For example, none of the Theseus data products used here include corrections for the Compton-Getting effect, a reference frame transformation effect leading to ENAs being over-counted in the ram direction (i.e., the direction of the Earth's motion around the sun) and under-counted in the antiram direction (i.e., opposite of the direction of motion). However, our real data illustrations using 2-degree Theseus maps generated by \citet{Osthus2022} and using official 6-degree maps generated by the IBEX Science Operations Center (ISOC) demonstrate good performance of the proposed methods, which will be applied to corrected Theseus maps and evaluated in detail in a series of follow-up papers in the heliospheric science literature. A slight modification of the ISOC map separations explored in this manuscript are being actively used by the space science community and appeared in the most recent ``official" IBEX mission data release in \citet{McComas2024}, further supporting the quality of these separations. \\
\indent We acknowledge that statistical solutions proposed in this paper are somewhat algorithmic or ad hoc in nature; for example, there is no coherent data generation model underlying our ribbon separation strategy. Given the difficulty of this problem and the inherent lack of separation identifiability in absence of concrete modeling assumptions, we found such an algorithmic approach appealing, as it allowed us to build physics assumptions and algorithmic guard rails directly into the separation procedure. When possible, we have justified each step of the proposed algorithms. Some approaches, however, were developed as a result of experimentation to find settings or criteria that ``worked" (i.e., they helped avoid poor separation performance for individual real data maps). In general, we make no claims of optimality of the proposed algorithms; rather, these choices ultimately came from extensive experience with these data and many, many iterations and tweaks to the algorithm, informed by close collaboration with heliospheric scientists who are actively using our separations for heliospheric science. Our simulation and real data demonstrations, however, illustrate the strong performance of the proposed methods and provide a strong body of evidence in favor of using proposed methods over existing competitors moving forward. \\
\indent Our ultimate goal was to provide a concrete, quantifiable advance in the quality of the IBEX source-specific maps used for subsequent scientific research into the complicated physics operating at the edges of our solar system. In service of this goal, this work illustrates the potential for statistical thinking to substantially contribute to scientific understanding; here, the same data inputs combined with more sophisticated physics-informed modeling and statistical thinking may provide heliospheric scientists better insights into ribbon and GDF dynamics than ever before. As more complicated and nuanced scientific data become available through advances in technology and computing, partnerships between data experts and scientific domain experts will be hugely important to facilitate scientific discovery.

\section{Acknowledgments}
This work was partially funded by the Laboratory Directed Research and Development (LDRD) Project 20220107DR. Dr. Beesley was also funded by the LDRD Richard Feynman Postdoctoral Fellowship 20210761PRD1. This work is approved for distribution under LA-UR-26-23882.

\section{Data Availability Statement}
The raw data used to generate the input maps will not be made publicly available. However, all simulated and real data inputs and outputs maps associated with this manuscript have been made publicly available at \url{https://github.com/lanl/IBEX_RibbonSeparation}.

\bibliographystyle{apalike}
\bibliography{Bib}

\end{document}